\documentclass[aps,pra,preprint,superscriptaddress,showpacs,floatfix]{revtex4}

\usepackage[dvips]{graphicx}

\newcommand{\bqn}{\begin{equation}}
\newcommand{\eqn}{\end{equation}}
\newcommand{\bqna}{\begin{eqnarray}}
\newcommand{\eqna}{\end{eqnarray}}
\newcommand{\bary}{\begin{array}{clcr}}
\newcommand{\eary}{\end{array}}

\begin{document}
\title
{Effects of laser polarization on photoelectron angular \\
distribution  through laser-induced continuum structure
}

\author{Gabriela Buica} \thanks{
permanent address: Institute for Space Sciences, P.O. Box MG-23, Ro 77125,
Bucharest-M\u{a}gurele, Romania}

\affiliation
{Institute of Advanced Energy, Kyoto University, 
Gokasho, Uji, Kyoto 611-0011, Japan}

\author{Takashi Nakajima}
\email[e-mail:]{t-nakajima@iae.kyoto-u.ac.jp}

\affiliation
{Institute of Advanced Energy, Kyoto University, 
Gokasho, Uji, Kyoto 611-0011, Japan}
\affiliation
{Institute for Solid State Physics, The University of Tokyo\\
  5-1-5 Kashiwanoha, Kashiwa, Chiba 277-8581, Japan}

\date{\today}
\begin{abstract}
We theoretically investigate the effects of laser polarization on 
photoelectron angular distribution through laser-induced continuum structure. 
We focus on a polarization geometry where the probe and dressing lasers 
are both linearly polarized, and change the relative polarization angle 
between them. 
We find that the total ionization yield and the branching ratio into different 
ionization channels change as a function of the relative polarization angle, 
and accordingly the photoelectron angular distribution is altered. 
We present specific results for the  $4p_{1/2}$-$6p_{1/2}$ and 
$4p_{3/2}$-$6p_{3/2}$ systems of the K atom, and show that the change 
of the polarization angle leads to the significant modification of
photoelectron angular distribution. 
\end{abstract}


\maketitle



\section{Introduction}

The similarity between the autoionizing structure (AIS)
and  the laser induced continuum structure (LICS)  is very well known. 
In an AIS process a discrete state lying above the ionization 
threshold is coupled to the continuum through configuration interaction, 
a resonance structure being created. 
In LICS  two bound states are coupled to the common continuum through two laser
fields (the probe and dressing lasers). The bound state coupled to the
continuum through a strong laser (dressing laser) can also create a 
resonance structure having AIS-like properties, but compared to the 
AIS resonance its position and width are now controllable by the 
frequencies and intensities of the lasers. 
 The first experimental observation of LICS was successfully reported in 
Refs. \cite{shao,cavalieri1}.
 More comprehensive information on LICS can
 be found in a review paper by Knight \textit{et al} \cite{knight}.

Through LICS not only the ionization yield but also a number of some other processes 
can be altered:
Several works based on LICS  investigated non-linear optical effects such as
the enhancement of third-high harmonic generation \cite{chara}.
In Ref. \cite{tak2}, effects of LICS on spin-polarization were studied for heavy 
alkali atoms. 
Recently, LICS for multiple continua was experimentally and theoretically investigated 
 \cite{yats}. 
In Ref. \cite{chen},  the control of ionization products in LICS was suggested 
for the case of decay into multiple continua.

It is well known that the photoelectron angular distribution (PAD) provides
more information about the ionization process than the angle-integrated 
ionization signal \cite{lamb}. 
PAD's of Na by the two linearly polarized lasers with a variable polarization 
were reported in Ref. \cite{dunc}, and the phase difference between the 
continua with same parity was extracted. 
By measuring PAD's of an alkali atom in a bichromatic laser field, 
a theoretical method was proposed in Ref. \cite{tak1} in order to extract 
the phase difference of the continua with opposite parities.   

Most recently we have theoretically investigated how LICS affects PAD \cite{takgab},
and specific results have been presented for the K atom. 
In Ref. \cite{takgab}, however, we have assumed that the probe and dressing lasers 
are linearly polarized along the same direction. A natural question would be 
how PAD is modified, through LICS, by changing the relative polarization angle 
between the probe and dressing lasers.

The aim of the present paper is to generalize our previous work \cite{takgab}.
We now vary the relative polarization angle and analyze the modifications of 
the ionization yield and the photoelectron angular distribution through LICS, 
and see how the enhancement or suppression of a particular ionization channel 
takes place. Since the photoelectrons ejected into different involved continua 
have different angular distributions, and those angular distributions depend 
on the relative polarization angle, we expect important modifications in terms of 
the ionization yields, branching ratios, and photoelectron angular distributions. 

The paper is organized as follows. 
In Sec. II we present the theoretical model:   The time-dependent amplitude 
equations which describe the dynamics of the LICS process are derived, 
and the ionization yield and photoelectron angular distribution are calculated. 
The consistency of our results has been checked using an alternative approach 
based on  the density matrix equations. 
The theoretical results obtained using these two formalisms are of course identical.
Further details about the density matrix formalism are provided at the beginning 
of Sec. III, and in  Appendix \ref{A}. 
Sec. III is mainly devoted to the discussions on the numerical results for the 
total and partial ionization rates, branching ratios, and PAD's at different 
polarization angles.
The atomic parameters needed for the LICS calculation are given in 
Appendices \ref{B}-\ref{D}.

\section{Theory}

The system we consider in this paper consists of an initially occupied 
$4p$ state, initially unoccupied $6p$ state, and the continuum states 
of the K atom together with the linearly polarized probe and dressing lasers 
that couple $4p$ and $6p$ to the continuum states, respectively. 
This implies that, prior to the interaction of the system with the 
probe and dressing lasers, K atoms in the ground $4s$ state have been 
excited to the $4p$ state by a linearly polarized auxiliary laser. 
By choosing an appropriate frequency of the auxiliary laser, we can 
selectively excite either $4p_{1/2}$ ($m_{j} = \pm 1/2$) or 
$4p_{3/2}$ ($m_{j} = \pm 1/2$), which will serve as an initial state 
in this work. For simplicity, we assume that the polarization axis of 
the dressing laser is parallel to that of the auxiliary laser, while  
the polarization axis of the probe laser can be arbitrary. 

Here we are interested in a particular geometry where the polarization vector 
of the dressing and auxiliary lasers are along the $z-$axis and 
that of the probe laser is assumed to lie in the $xz$-plane, as shown 
in Fig. \ref{fig1}.
For such a case, the polarization vector of the probe laser is defined as
$\textbf{e}^{(p)}= \textbf{e}_{1}\sin\theta_{p}+\textbf{e}_{3}\cos\theta_{p}$, where
$\textbf{e}_{1}$, and $ \textbf{e}_{3}$ are the unit vectors along the $x$ and $z$ axes,
respectively, and $\theta_{p}$ represents the polarization angle of the probe laser 
with respect to that of the dressing laser.
Defining the frequencies of the probe and dressing lasers as $\omega_{p}$ and 
$\omega_{d}$, respectively, the total electric field vector can be written as, 
\bqn \textbf {E}(t) =\sum_{\alpha=p,d}{\cal E}_{0\alpha}(t) 
 \textbf {e}^{(\alpha)} \cos(\omega_{\alpha}t ).
 \label{field}
\eqn

\noindent
A Gaussian temporal profile was employed for the amplitude of the laser fields:
${\cal E}_{0\alpha}(t)= {\cal E}_{0\alpha} 
\exp{ \left[-4 \ln2 \ \left(t/ \tau_{\alpha}\right)^2\right]}$, 
where $\tau_{\alpha}$ represents the temporal width for the full width at 
half maximum (FWHM) of the probe or dressing pulse 
with $\alpha = p$ or $d$, indicating the probe and the dressing pulses. 
$\textbf {e}^{(\alpha)}$  is the polarization vector of the laser pulse $\alpha$.

Based on the above assumptions, the level scheme we consider in this paper
is now described in Figs. \ref{fig2}(a)-\ref{fig2}(c) for the 
K $4p_{1/2}$-$6p_{1/2}$ system, at 
$\theta_p=0^\circ$, $\theta_p=90^\circ$, and in between, 
i.e., $0^\circ< \theta_p<90^\circ$. 
If both polarization axes of the probe and dressing lasers are parallel, 
i.e.  $\theta_p=0^\circ$ as shown in Fig. \ref{fig2}(a), due to the selection rule 
$m_{j'}= m_{j}$ (where the prime index is used for quantum numbers of the continuum) 
the {\it entire} $4p_{1/2}$-$6p_{1/2}$ system with $m_{j} = \pm 1/2$ can be decomposed 
into the {\it two independent subsystems} which consist of 
$4p_{1/2}$ ($m_{j} =\pm 1/2$), $6p_{1/2}$ ($m_{j} =\pm 1/2$), and 
the continua $\epsilon s$ ($m_{j'} =\pm 1/2$) and $\epsilon d$ ($m_{j'} =\pm 1/2$). 
The ionization yields for these subsystems are obviously symmetric to each other, 
and for simplicity we can consider only one of them, as already explained in our 
previous paper \cite{takgab}. 
\noindent
Similarly, at $\theta_p=90^\circ$, because of the selection rule $m_{j'}= m_{j} \pm 1$, 
the {\it entire} $4p_{1/2}$-$6p_{1/2}$ system with $m_{j} = \pm 1/2$ can be 
decomposed into the {\it two independent subsystems} consisting of 
$4p_{1/2}$ ($m_{j} = \pm 1/2$), $6p_{1/2}$ ($m_{j} = \mp 1/2$), and the continua 
$\epsilon s$ ($m_{j'} =\mp 1/2$) and $\epsilon d$ ($m_{j'} =\mp 1/2$) with additional
incoherent channels $\epsilon d$ ($m_{j'} =\pm 3/2$), as shown in  Fig. \ref{fig2}(b). 
Again, both subsystems are completely symmetric, and it is sufficient to study 
only one of the two subsystems. 
\noindent   
However, for the intermediate values of the polarization angle, 
$0^\circ< \theta_p<90^\circ$ as shown in Fig. \ref{fig2}(c), because of the 
selection rules  $m_{j'}= m_{j} \pm 1$ (from the perpendicular component 
of the probe polarization vector with respect to the quantization axis, 
$\theta_p=90^\circ$) and $m_{j'}= m_{j}$ (from the parallel 
component, $\theta_p=0^\circ$), the {\it entire} system $4p_{1/2}$-$6p_{1/2}$  
{\it cannot} be decomposed into the two independent subsystems anymore, 
and the entire system, $4p_{1/2}$, $6p_{1/2}$, and the continua 
with all possible magnetic sublevels, have to be taken into account 
at the same time. 
\noindent
The continuum states  $|c_b\rangle$, $(b=5, 8)$, not presented in Fig. \ref{fig2},
have the same quantum numbers as  the continuum states 
$|c_a\rangle$, ($a=1, 4$, for $ m_{j'} =+ 3/2,+1/2, -1/2,-3/2  $), 
but they correspond  to a different value of energy because of the incoherent 
one-photon ionization from  $6p_{1/2}$ by the probe laser.
A similar level scheme, taking into account appropriate dipole selection rules,
has been considered for the K $4p_{3/2}$-$6p_{3/2}$ system. 

In order to observe LICS, it is necessary that the initially occupied 
$4p_{1/2}$ (or $4p_{3/2}$) state (denoted as $|1\rangle $ for the magnetic sublevel 
having  $m_{j} = + 1/2$ and $|2\rangle $ for the magnetic sublevel having $m_{j} = - 1/2$),  
and the initially unoccupied $6p_{1/2}$ (or $6p_{3/2}$) state (denoted as $|3\rangle $ 
for  $m_{j} = + 1/2$ and $|4\rangle $ for  $m_{j} = - 1/2$) are coupled by the 
probe and dressing lasers whose frequencies nearly satisfy the two-photon resonance, 
i.e., $E_{4p}+\omega_{p} \simeq E_{6p}+\omega_{d}$. 
As long as we use a ns laser with appropriate intensities and detunings, 
it is perfectly valid to treat each $4p_{1/2}-6p_{1/2}$ and $4p_{3/2}-6p_{3/2}$ system 
separately, as it was explained in our previous paper \cite{takgab}. 

It should be mentioned that we expect a different behavior of the two systems: 
For the K $4p_{1/2}-6p_{1/2}$ system the initial state, $4p_{1/2}$, is an isotropic 
mixture of  all  possible magnetic sublevels (recall that the magnetic sublevels 
$m_j=\pm 1/2$  are equally populated by the auxiliary laser), implying that the 
initial state is spherically symmetric. 
It is obvious that the PAD from the spherically symmetric initial state orientates 
along the polarization axis of the probe laser if the dressing laser is off. 
PAD changes neither its shape nor magnitude \cite{sob}. 
That is not the case for the K $4p_{3/2}-6p_{3/2}$ system, since not all possible 
magnetic sublevels are excited with the same probability, and accordingly the initial state, 
$4p_{3/2}$, is non-spherical (polarized). 
Therefore, we expect quite different  modification of PAD for the  $4p_{1/2}-6p_{1/2}$ 
and $4p_{3/2}-6p_{3/2}$ systems through LICS, by varying the relative polarization angle, 
as we have already seen for $ \theta_p=0^\circ$ \cite{takgab}.
Further details will be provided in Sec. III.

\subsection{Time-dependent amplitude equations}

In order to study the temporal evolution of the atomic system in laser field, we have 
used the standard procedure as described in our previous paper  \cite{takgab}. 
Briefly, we solve the following set of time-dependent amplitude equations: 
\bqna
&&\dot{u}_{1} =   - \frac{1}{2} \widetilde{\Gamma}_1  u_{1} 
            -i \Omega_{13} \left( 1 - \frac{i}{q_{13}} \right) u_{3} 
	    -i \Omega_{14} \left( 1 - \frac{i}{q_{14}} \right) u_{4}  \;,
\label{amp1} \\
&&\dot{u}_{2} =   - \frac{1}{2} \widetilde{\Gamma}_2  u_{2} 
            -i \Omega_{23} \left( 1 - \frac{i}{q_{23}} \right) u_{3} 
	    -i \Omega_{24} \left( 1 - \frac{i}{q_{24}} \right) u_{4}  \;,
\label{amp2} \\
&&\dot{u}_{3} = \left( i  \delta - \frac{1}{2} \widetilde{\Gamma}_3
             \right) u_{3}
     - i \Omega_{31} \left( 1 - \frac{i}{q_{31}} \right) u_{1} 
     - i \Omega_{32} \left( 1 - \frac{i}{q_{32}} \right) u_{2}  \;,
\label{amp3}\\
&&\dot{u}_{4} = \left( i  \delta - \frac{1}{2} \widetilde{\Gamma}_4
             \right) u_{4}
     - i \Omega_{41} \left( 1 - \frac{i}{q_{41}} \right) u_{1} 
     - i \Omega_{42} \left( 1 - \frac{i}{q_{42}} \right) u_{2}  \;,
\label{amp4}
\eqna
\noindent
where $u_{j}$'s ($j={1,4}$) represent the
probability amplitudes of states $| j \rangle$. Note that all the probability amplitudes 
for the continuum states have already been adiabatically eliminated in the 
Eqs. (\ref{amp1})-{(\ref{amp4}). 
$\delta$ is a two-photon detuning defined by $\delta= \delta_{static} +
\delta_{stark}$,
 where the static detuning is defined by
$\delta_{static} = ( E_{1} + \hbar \omega_{p} ) - ( E_{3} + \hbar \omega_{d} )$,
 and $\delta_{stark}$ is a total dynamic ac Stark shift defined by
$\delta_{stark}=(S_{1}^{(p)} + S_{1}^{(d)}) - (S_{3}^{(p)} + S_{3}^{(d)})$. 
In all the numerical results presented in this work the zero point of the
detuning has been chosen  such that 
$\delta \to \delta -\delta_{stark}^{max}$,  
since the ac Stark shifts simply translate the LICS resonance on the detuning scale.
The superscript of $\delta_{stark}^{max}$ means  that the ac Stark shift 
is calculated at the peak value of laser intensity.
$D_{jc}^{(\alpha)}$'s 
are the bound-free matrix elements by laser $\alpha$
($\alpha = p$ or $d$) from the bound state $| j \rangle$ to the continuum 
$| c \rangle$, which are connected to the partial ionization widths 
through the relation 
$\Gamma_{jc}^{(\alpha)} = 2 \pi |D_{jc}^{(\alpha)}|^{2}$. 
$\widetilde{\Gamma}_{j}$  represents the total ionization width of
state $| j \rangle$, i.e., 
$\widetilde{\Gamma}_{j} \equiv \gamma_{j} + \Gamma_{j}^{(p)}$ (for $j=1, 2$)
and
$\widetilde{\Gamma}_{j} \equiv \gamma_{j} + \Gamma_{j}^{(d)}+ \Gamma_{j}^{(p)}
$ (for $j=3, 4$), where 
$\gamma_{j}$  is the phenomenologically introduced spontaneous 
decay width of state $| j \rangle$.
In the above equations the two-photon Rabi frequency, $\Omega_{ij} $,  can be 
written as a sum of the partial two-photon Rabi frequencies into the coherent 
$\epsilon s$ and $\epsilon d$ continua of energy $\epsilon $, 
i.e.,   
\bqn
\Omega _{ij}\left( 1 - \frac{i}{q_{ij}} \right)
= \sum_{\beta = \epsilon s, \epsilon d} 
\Omega_{ij}^{\beta} \left( 1 - \frac{i}{q_{ij}^{\beta}} \right)  \;,
\label{omegaandq}
\eqn
where $q_{ij}$ and $q_{ij}^{\beta}$ represent the total and partial asymmetry parameters, 
respectively.
It is very well known that, for the light alkali-metals such as Li, N, and K, 
the dependence of  radial matrix elements and phase shifts on the total angular momentum 
quantum number $j$ is very small and it can be neglected \cite{lambls}. 
Now, the following relations are satisfied by the angle-integrated atomic parameters:
$ \Gamma_{1}^{(\alpha)}= \Gamma_{2}^{(\alpha)}$, 
$ \Gamma_{3}^{(\alpha)}= \Gamma_{4}^{(\alpha)}$,
$ S_{1}^{(\alpha)}= S_{2}^{(\alpha)}$, $ S_{3}^{(\alpha)}= S_{4}^{(\alpha)}$, 
$\Omega_{13} = \Omega_{24}$,  $\Omega_{14} = -\Omega_{23}$,  $q_{13} = q_{24}$,
and $q_{14} = -q_{23}$. 
Details about the calculation of the atomic parameters such as Rabi frequencies,
ionization widths, ac Stark shifts and asymmetry parameters are given in 
the Appendices \ref{B}-\ref{D}.

Since the behavior of the population dynamics in the continuum is of our interest, 
we also need the following set of amplitude equations for the continua:
\bqna
&&\dot{u}_{c_a} = - i \delta_{c_a} u_{c_a} 
- i D_{c_a1}^{(p)} u_{1} - i D_{c_a2}^{(p)} u_{2}
- i D_{c_a3}^{(d)} u_{3} - i D_{c_a4}^{(d)} u_{4}
\;,
\label{ampcoh_a}\\
&&\dot{u}_{c_b} = - i \delta_{c_b} u_{c_b}
- i D_{c_b3}^{(p)} u_{3} - i D_{c_b4}^{(p)} u_{4}
\;.
\label{ampcoh_b}
\eqna
Here ${u}_{c_a}$ represents the probability amplitude of the coherent continuum state 
$|c_{a}\rangle$  ($a=1,4$), and ${u}_{c_b}$ represents the probability amplitude  
of the incoherent continuum state $|c_{b}\rangle$ ($b=5,8$).
As already explained at the beginning of Sec. II  the incoherent continuum states 
$|c_{b}\rangle$ are not presented in Fig. \ref{fig2} to avoid the complexity of the figure.  
They have the same quantum numbers as the coherent continuum states $|c_{a}\rangle$, 
but located at different energies. 

Having solved Eqs. (\ref{amp1})-{(\ref{amp4}) and Eqs. (\ref{ampcoh_a}-\ref{ampcoh_b}), 
we can now calculate the total (angle-integrated) ionization yield, $R(t)$, 
from the relation, 
\bqn
R(t) =  \sum_{a=1}^{8} R_{c_{a}}(t), \label{totalyield}
\eqn
where the partial photoelectron yields $R_{c_{a}}(t)$, into each coherent and incoherent 
continuum state $|c_{a}\rangle$, ($a=1, 8$) can be calculated through the following formulae:
\bqna
R_{c_{1}}(t) &=&\int_{-\infty}^{t}dt'   \Gamma_{1c_{1}}^{(p)}  |u_1|^2,
 \label{partialc1} \\
R_{c_{2}}(t) &=&\int_{-\infty}^{t}dt'\left\{ 
 \Gamma_{1c_{2}}^{(p)}  |u_1|^2 + 
 \Gamma_{2c_{2}}^{(p)}  |u_2|^2 + \Gamma_{3c_{2}}^{(d)}  |u_3|^2 
 \right. \nonumber\\ &+& \left.
4 \;{\rm Im}
{\left[\Omega_{13}^{c_{2}}\left(1+\frac{i}{q_{13}^{c_{2}}}\right)\right]}
 {\rm Re}(u_1u_3^*)
+4\; {\rm Im}
{\left[\Omega_{23}^{c_{2}}\left(1+\frac{i}{q_{23}^{c_{2}}}\right)\right]}
 {\rm Re}(u_2u_3^*) 
 \right\},
 \label{partialc2} \\
R_{c_{3}}(t) &=&\int_{-\infty}^{t}dt'\left\{ 
 \Gamma_{1c_{3}}^{(p)}  |u_1|^2 + 
 \Gamma_{2c_{3}}^{(p)}  |u_2|^2 + \Gamma_{4c_{3}}^{(d)}  |u_4|^2 
 \right. \nonumber\\ &+& \left.
4 \;{\rm Im}
{\left[\Omega_{14}^{c_{3}}\left(1+\frac{i}{q_{14}^{c_{3}}}\right)\right]}
 {\rm Re}(u_1u_4^*)
+4\; {\rm Im}
{\left[\Omega_{24}^{c_{3}}\left(1+\frac{i}{q_{24}^{c_{3}}}\right)\right]}
 {\rm Re}(u_2u_4^*) 
 \right\},  \label{partialc3}\\
 R_{c_{4}}(t) &=&\int_{-\infty}^{t}dt' 
   \Gamma_{2c_{4}}^{(p)}  |u_2|^2,  \label{partialc4}\\
R_{c_{5}}(t) &=&\int_{-\infty}^{t}dt' 
   \Gamma_{3c_{5}}^{(p)}  |u_3|^2,  \label{partialc5}\\
 R_{c_{6}}(t) &=&\int_{-\infty}^{t}dt' 
   \left[ \Gamma_{3c_{6}}^{(p)}  |u_3|^2 +
    \Gamma_{4c_{6}}^{(p)}  |u_4|^2\right], \label{partialc6}\\ 
 R_{c_{7}}(t) &=&\int_{-\infty}^{t}dt' 
   \left[ \Gamma_{3c_{7}}^{(p)}  |u_3|^2 +
    \Gamma_{4c_{7}}^{(p)}  |u_4|^2\right], \label{partialc7}\\    
 R_{c_{8}}(t) &=&\int_{-\infty}^{t}dt' 
   \Gamma_{4c_{8}}^{(p)}  |u_4|^2.  \label{partialc8}        
\eqna

\noindent
Since the total ionization yield is a sum of ionization into the coherent 
and incoherent continua, it might be rewritten as, 
\bqn 
R(t) =  \sum_{\beta = \epsilon s, \epsilon d} 
\left[
\sum_{a=1}^{4} R_{c_{a}}^{\beta}(t) +\sum_{b=5}^{8} R_{c_{b}}^{\beta}(t)
\right]	=
\sum_{\beta = \epsilon s, \epsilon d} 
\left[ R_{\beta}(t) +R_{\beta}^{incoh}(t) \right].
\label{ratesd}
\eqn
To see the effects of LICS, it is useful to calculate the branching ratio, $B$,
 defined as the ratio between the partial
ionization yield into each coherent continuum $\epsilon d$ and $\epsilon s$:
\bqn
B=\frac{R_{\epsilon d}}{R_{\epsilon s}}.
\label{branching}
\eqn
The total ionization yield given by Eq. (\ref{totalyield}) and the partial 
ionization yields given by Eqs. (\ref{partialc1})-(\ref{partialc8}) are calculated 
at the end of the pulses. 

\subsection{Photoelectron angular distribution}

For the purpose of calculating  photoelectron angular distribution we need equations before 
angle integration. 
In order to simplify the calculation of  the bound-free dipole matrix elements 
we use a partial wave expansion for the continuum of an alkali-metal atom in a coupled 
$|(l's')j'm_{j'}\rangle$  basis: 
\bqn
|\textbf{k} ; m_{s'} \rangle   
 = \sum_{l' , m_{l'} , j'} a_{l' m_{l'}} (-1)^{l' - 1/2 + m_{l'} + m_{s'}}
\sqrt{2 j' + 1} 
\left(\begin{array}{clcr}
          l'       &     1/2    &          j'        \\
         m_{l'}    &     m_{s'}  &        -m_{j'} \\
        \end{array}\right)
| \textbf{{k}}; (l's') j' m_{j'} \rangle
\;,
\eqn

\noindent
where $\textbf{k}$ represents the wave vector of  photoelectron, 
$a_{l' m_{l'}} = 4 \pi i^{l'} e^{-i \delta_{l'}} Y_{l' m_{l'}}(\Theta,\Phi)$,  
and $\delta_{l'}$ is the phase shift which is a sum of the Coulomb phase shift 
and the scattering phase shift; recall that the prime index indicates 
a quantum number for the continuum state. 

We are interested in PAD as a function of polarization angle $\theta_{p}$. 
If the final spin state of the photoelectron is not detected, we have to 
incoherently sum over the final spin projection, $m_{s'}$. 
The partial photoelectron yield into a solid angle, $\Omega_{\textbf{k}}$, defined 
by the polar angle $\Theta$, and the azimuthal angle $\Phi$, can be written as, 
\noindent
\bqna 
\left. \frac{dR (\Theta,\Phi)}{dt d\Omega_{\textbf{k}}}  \right\vert_{m_j=\pm 1/2} = 
0.589 {\pi}\sum_{m_{s'}=\pm 1/2}
&& \left[
   \left|\sum_{j=1}^{2}
       \sqrt{\Gamma_{j}^{(p,m_{s'})}(\Theta,\Phi)} \;u_{j} 
   +\sum_{j=3}^{4}
       \sqrt{\Gamma_{j}^{(d,m_{s'})}(\Theta,\Phi)} \;u_{j}                  
   \right|^2 
\right. \nonumber\\ &+& \left.    
   \left|\sum_{j=3}^{4}
       \sqrt{\Gamma_{j}^{(p,m_{s'})}(\Theta,\Phi)}\;  u_{j} 
   \right|^2   
\right] 
\;,
\label{pad}
\eqna
\noindent
where $0.589 \pi$ is a conversion factor for the appropriate normalization, 
and the formula is valid for both $4p_{1/2}-6p_{1/2}$ and
$4p_{3/2}-6p_{3/2}$ systems, for $m_j= \pm 1/2$ of the initial state.
\noindent
After the considerable angular momentum algebra,  we obtain expressions for the
differential ionization widths $\Gamma_{i}^{(\alpha,m_{s'})}(\Theta,\Phi)$  
from states $|1\rangle$ and $|3\rangle$ for the $4p_{1/2}-6p_{1/2}$ system:   
\bqna
\Gamma_{j}^{(\alpha,+1/2)}(\Theta,\Phi) &=& 
\left| 
   - \frac{1}{3} R_{j\epsilon s}^{(\alpha)} e^{i \delta_{s}} 
     Y_{00}(\Theta,\Phi) e_0^{(\alpha)} 
     +\frac{2}{3\sqrt{5}}R_{j \epsilon d}^{(\alpha)} e^{i \delta_{d}} 
     Y_{20}(\Theta,\Phi) e_0^{(\alpha)}    
     \right. \nonumber \\ 
&+ & \left.
 \frac{1}{\sqrt{15}}  R_{j\epsilon d}^{(\alpha)} e^{i \delta_{d}} 
     Y_{21}(\Theta,\Phi)  e_1^{(\alpha)} 
+
 \frac{1}{\sqrt{15}}  R_{j\epsilon d}^{(\alpha)} e^{i \delta_{d}} 
    Y_{2-1}(\Theta,\Phi) e_{-1}^{(\alpha)}         
\right|^2 
I_{\alpha}
\;,
\label{pad12gamma13p}
\\
\Gamma_{j}^{(\alpha,-1/2)}(\Theta,\Phi) &=&
\left| 
   - \frac{\sqrt{2}}{3} R_{j\epsilon s}^{(\alpha)} e^{i \delta_{s}} 
     Y_{00}(\Theta,\Phi) e_{-1}^{(\alpha)} 
     -\frac{\sqrt{2}}{3\sqrt{5}}R_{j \epsilon d}^{(\alpha)} e^{i \delta_{d}} 
     Y_{20}(\Theta,\Phi) e_{-1}^{(\alpha)}              \right.
\nonumber \\ 
&-& 
 \frac{2}{\sqrt{15}}  R_{j \epsilon d}^{(\alpha)} e^{i \delta_{d}} 
     Y_{22}(\Theta,\Phi)  e_1^{(\alpha)} 
-
\left. \frac{\sqrt{2}}{\sqrt{15}}  R_{j \epsilon d}^{(\alpha)} e^{i
\delta_{d}} 
    Y_{21}(\Theta,\Phi) e_{0}^{(\alpha)}         
\right|^2 
I_{\alpha}
\;,
\label{pad12gamma13m}
\eqna
\noindent
and for the differential ionization widths 
$\Gamma_{j}^{(\alpha,m_{s'})}(\Theta,\Phi)$, from states $|2\rangle$ and $ |4\rangle$: 
\bqna
\Gamma_{j}^{(\alpha,+1/2)}(\Theta,\Phi) &=& 
\left| 
    \frac{\sqrt{2}}{3} R_{j\epsilon s}^{(\alpha)} e^{i \delta_{s}} 
     Y_{00}(\Theta,\Phi) e_1^{(\alpha)} 
     +\frac{\sqrt{2}}{3\sqrt{5}}R_{j \epsilon d}^{(\alpha)} e^{i \delta_{d}} 
     Y_{20}(\Theta,\Phi) e_1^{(\alpha)}   \right.           
	\nonumber \\ 
&+ &\left.
\frac{\sqrt{2}}{\sqrt{15}}  R_{j\epsilon d}^{(\alpha)} e^{i
\delta_{d}} 
     Y_{2-1}(\Theta,\Phi)  e_0^{(\alpha)}  
 +
 \frac{2}{\sqrt{15}}  R_{j\epsilon d}^{(\alpha)} e^{i \delta_{d}} 
    Y_{2-2}(\Theta,\Phi) e_{-1}^{(\alpha)}         
\right|^2
I_{\alpha}
\;,
\label{pad12gamma24p}
\\
\Gamma_{j}^{(\alpha,-1/2)}(\Theta,\Phi) &=&
\left| 
     \frac{1}{3} R_{j\epsilon s}^{(\alpha)} e^{i \delta_{s}} 
     Y_{00}(\Theta,\Phi) e_{0}^{(\alpha)} 
     -\frac{2}{3\sqrt{5}}R_{j \epsilon d}^{(\alpha)} e^{i \delta_{d}} 
     Y_{20}(\Theta,\Phi) e_{0}^{(\alpha)}              
\right. \nonumber \\ 
&-&\left.
 \frac{1}{\sqrt{15}}  R_{j\epsilon d}^{(\alpha)} e^{i \delta_{d}} 
     Y_{21}(\Theta,\Phi)  e_1^{(\alpha)} 
-
 \frac{1}{\sqrt{15}}  R_{j\epsilon d}^{(\alpha)} e^{i
\delta_{d}} 
    Y_{2-1}(\Theta,\Phi) e_{-1}^{(\alpha)}         
\right|^2 
I_{\alpha}
\;,
\label{pad12gamma24m}
\eqna
where $R_{j \epsilon s}^{(\alpha)}$ and $R_{j \epsilon d}^{(\alpha)}$ represent 
the radial bound-free matrix elements from state $|j\rangle$ ( $j=1,4$) 
to the continua $\epsilon s$ and $\epsilon d$, respectively,   
by laser $\alpha$ ($\alpha = p$ or $d$),  evaluated in atomic units.
Here $e^{(\alpha)}_q$, with $q=0,\pm 1$, are the spherical components of the polarization
vector of laser $\alpha$, namely $e^{(\alpha)}_0= \cos \theta_{\alpha}$, and 
$e^{(\alpha)}_{\pm 1}= \mp \sin \theta_{\alpha}/\sqrt{2}$.  
The laser intensities, $I_{p}$ and $I_{d}$, are expressed in W/cm$^{2}$. 
For the coherent continuum, the relevant phase shifts are $\delta_{s}=1.937$ 
and $\delta_{d}=-6.574$, which are the sum of the Coulomb phase shifts, 
$\delta_{s}^{C}=-4.924$ and $\delta_{d}^{C}=-7.551$, and the scattering 
phase shifts, $\pi \mu_{s}= 6.861$ and $\pi \mu_{d}= 0.977$ with 
$\mu_{l}$ ($l=s,d$) being the quantum defects estimated from the 
linear extrapolation of the bound Rydberg $s$ and $d$ series of the K atom 
to the continuum energy of interest.    
Eq. (\ref{pad}) together with Eqs. (\ref{pad12gamma13p})-(\ref{pad12gamma24m})
gives PAD for the $4p_{1/2}-6p_{1/2}$ system with appropriate normalization,
so that the angle-integrated quantity becomes identical to the 
total ionization yield calculated with Eq. (\ref{totalyield}).

Similarly the differential ionization widths from states $|1\rangle $ and $|3\rangle$ 
for the $4p_{3/2}-6p_{3/2}$ system, are given by, 
\bqna
\Gamma_{j}^{(\alpha,+1/2)}(\Theta,\Phi) &=&
\left| 
     \frac{\sqrt{2}}{3} R_{j\epsilon s}^{(\alpha)} e^{i \delta_{s}} 
     Y_{00}(\Theta,\Phi) e_{0}^{(\alpha)}     
   -\frac{2\sqrt{2}}{3\sqrt{5}}   R_{j \epsilon d}^{(\alpha)} e^{i\delta_{d}}         
Y_{20}(\Theta,\Phi)  e_{0}^{(\alpha)}    \right.
\nonumber \\ &-&
 \left.     
\frac{\sqrt{2}}{\sqrt{15}}  R_{j \epsilon d}^{(\alpha)} e^{i \delta_{d}}
  Y_{21}(\Theta,\Phi)e_{1}^{(\alpha)}   
  -  \frac{\sqrt{2}}{\sqrt{15}}  R_{j \epsilon d}^{(\alpha)} e^{i \delta_{d}}
   Y_{2-1} (\Theta,\Phi) e_{-1}^{(\alpha)}   
      \right|^2 
I_{\alpha} ,\label{pad32gamma13p} 
\\
\Gamma_{j}^{(\alpha,-1/2)}(\Theta,\Phi)  &=&
\left|-
 \frac{1}{3} R_{j\epsilon s}^{(\alpha)} e^{i \delta_{s}} 
     Y_{00}(\Theta,\Phi) e_{-1}^{(\alpha)}     
    - \frac{1}{3\sqrt{5}}  R_{j \epsilon d}^{(\alpha)} e^{i \delta_{d}}
  Y_{20}(\Theta,\Phi)e_{-1}^{(\alpha)} \right.
\nonumber \\ &-&
 \left. 
   \frac{\sqrt{2}}{\sqrt{15}}  R_{j \epsilon d}^{(\alpha)} e^{i \delta_{d}}
   Y_{22} (\Theta,\Phi) e_{1}^{(\alpha)} 
   - \frac{1}{\sqrt{15}}   R_{j \epsilon d}^{(\alpha)} e^{i\delta_{d}}         
Y_{21}(\Theta,\Phi)  e_{0}^{(\alpha)}     
    \right|^2
    I_{\alpha}
\;,\label{pad32gamma13m} 
\eqna
and from states $|2\rangle $ and $|4\rangle$ they are derived as, 
\noindent
\bqna
\Gamma_{j}^{(\alpha,+1/2)}(\Theta,\Phi) &=&
\left| 
    - \frac{1}{3} R_{j\epsilon s}^{(\alpha)} e^{i \delta_{s}} 
     Y_{00}(\Theta,\Phi) e_{1}^{(\alpha)}     
   -\frac{1}{3\sqrt{5}}   R_{j \epsilon d}^{(\alpha)} e^{i\delta_{d}}         
	Y_{20}(\Theta,\Phi)  e_{1}^{(\alpha)}    \right.
\nonumber \\ &-&
 \left.      
\frac{1}{\sqrt{15}}  R_{j \epsilon d}^{(\alpha)} e^{i \delta_{d}}
  Y_{2-1}(\Theta,\Phi)  e_{0}^{(\alpha)}  
-\frac{\sqrt{2}}{\sqrt{15}}  R_{j \epsilon d}^{(\alpha)} e^{i \delta_{d}}
   Y_{2-2} (\Theta,\Phi) e_{-1}^{(\alpha)}   
      \right|^2 
I_{\alpha}, \label{pad32gamma24p}  
\\
\Gamma_{j}^{(\alpha,-1/2)}(\Theta,\Phi)  &=&
\left|
 \frac{\sqrt{2}}{3} R_{j\epsilon s}^{(\alpha)} e^{i \delta_{s}} 
     Y_{00}(\Theta,\Phi) e_{0}^{(\alpha)}  
 -\frac{2\sqrt{2}}{3\sqrt{5}}   R_{j \epsilon d}^{(\alpha)} e^{i\delta_{d}}         
	Y_{20}(\Theta,\Phi)  e_{0}^{(\alpha)}       \right.
\nonumber \\ &-&
 \left.      
\frac{\sqrt{2}}{\sqrt{15}}  R_{j \epsilon d}^{(\alpha)} e^{i \delta_{d}}
  Y_{21}(\Theta,\Phi)  e_{1}^{(\alpha)}     
 - \frac{\sqrt{2}}{\sqrt{15}}  R_{j \epsilon d}^{(\alpha)} e^{i \delta_{d}}
   Y_{2-1} (\Theta,\Phi) e_{-1}^{(\alpha)}       
    \right|^2     I_{\alpha}
\;.\label{pad32gamma24m}  
\eqna

\noindent
The Eqs. (\ref{pad12gamma13p})-(\ref{pad32gamma24m}) are applicable for both probe 
and dressing lasers. Recalling that the polarization vector of the dressing laser 
is parallel to the quantization axis ($\theta_d=0^\circ$), 
the relative polarization angle between the
probe and dressing lasers becomes identical to $\theta_p$. 
Because of the symmetry properties of spherical harmonics, we can show that
$\Gamma_{1}^{(\alpha,m_{s'})}$ and $\Gamma_{3}^{(\alpha,m_{s'})}$ is equal to
$\Gamma_{2}^{(\alpha,-m_{s'})}$ and $\Gamma_{4}^{(\alpha,-m_{s'})}$, respectively,  
by interchanging $e_{q}^{(\alpha)}$ and $e_{-q}^{(\alpha)}$. 

\section{Numerical Results and Discussion}

In this section we present numerical results and discussions. 
All the necessary single and effective two-photon dipole matrix elements needed 
for our schemes have been obtained using quantum defect theory and Green`s function technique. 
The calculated  atomic parameters  such as Rabi frequencies, asymmetry parameters, 
ac Stark shifts,  for the $4p_{1/2}-6p_{1/2}$ and $4p_{3/2}-6p_{3/2}$  systems, 
are  listed in Tables I and II, respectively.

For the K $4p_{1/2}-6p_{1/2}$ system only  Rabi frequencies  are $\theta_{p}$ dependent, 
ionization widths and ac Stark shifts by the probe laser do not present any $\theta_{p}$ 
dependence since the initial state is isotropic  \cite{sob,he} 
(recall that by using a linearly polarized auxiliary laser the ground state, $4s_{1/2}$, of the K atom is excited to the $4p_{1/2}$ state, thus all the magnetic  sublevels for the $4p_{1/2}$ state are occupied with the same probability).
In contrast, for the K $4p_{3/2}-6p_{3/2}$ system the atomic parameters such as  
Rabi frequencies, and  ionization widths and ac Stark shifts by the probe laser 
depend on the polarization angle  $\theta_{p}$. This is due to the fact that not 
all the magnetic  sublevels for the initial state, $4p_{3/2}$, are occupied with the same probability.
The total and partial asymmetry parameters are independent on the laser fields 
\cite{knight}, and, obviously, do not depend on the polarization angle of the probe 
laser for both K $4p_{1/2}-6p_{1/2}$ and $4p_{3/2}-6p_{3/2}$ systems. 

When we solve the set of amplitude equations a special care has to be taken:
Note that the probability amplitudes of the initially occupied states, $|1\rangle$ 
and $|2\rangle$, have {\it arbitrary phases}. 
However, if we use the amplitude equations, once the initial conditions are given for 
the probability amplitudes $u_i(t=-\infty)$, $i=1,4$, initial coherence inevitably exists 
due to nonzero value of  $u_iu^*_j$ (with $i \neq j$). 
 We have to avoid any coherent 
interference between the ionization paths starting from $|1\rangle$ and $|2\rangle$. 
Therefore, we should \textit{separately} solve the set of amplitude equations 
Eqs. (\ref{amp1})-(\ref{amp4}) with {\it either} $u_{1}(t=-\infty) =1 $ 
and $u_{i}(t=-\infty)=0$ (for $i=2,3,4$), \textit{or}  $u_{2}(t=-\infty) =1 $ 
and $u_{i}(t=-\infty)=0$ (for $i=1,3,4$), and average the photoelectron 
angular distribution given by Eq. (\ref{pad}) over $m_j$ of the initial state:
\bqn
\frac{dR (\Theta,\Phi)}{dt d\Omega_{\textbf{k}}}  = 
\frac{1}{2} \left[  \left.  \left.
 \frac{dR (\Theta,\Phi)}{dt d\Omega_{\textbf{k}}}  \right\vert_{m_j= + 1/2}
 +\frac{dR (\Theta,\Phi)}{dt d\Omega_{\textbf{k}}} \right\vert_{m_j= - 1/2} 
 \right].
\label{pad_total}  
\eqn

Pulse durations and peak laser intensities are chosen to be $\tau_{p} = 1$ ns (FWHM) 
and $I_{p} = 1 $ MW/cm$^{2}$ for the probe laser, and 
$10 $ ns $\leq \tau_{d} \leq  15$ ns (FWHM) and 
$100 $ MW/cm$^{2}$ $\leq I_{d} \leq  500$ MW/cm$^{2}$ for the  dressing laser.
If the probe and the dressing pulse durations are comparable the LICS resonance 
profile is going to be smeared out \cite{takgab} due to the ac Stark shifts. 
In order to circumvent this problem the pulse duration of the dressing laser 
was chosen to be much longer than that of the probe laser, since, under the 
condition that $\tau_d \gg \tau_p$, atomic states are quasi-statistically 
Stark-shifted by the strong dressing pulse during the interaction with the probe pulse.
By substituting the atomic parameters listed in Tables I and II into 
Eqs. (\ref{amp1})-(\ref{amp4}), 
we can easily solve those equations for the given peak intensities, detunings, and  
temporal profile of the lasers. 
Once the solution is obtained for $u_{i}(t)$ ($i=1,4$), the total and partial 
ionization yields  can be calculated from   Eq. (\ref{totalyield})  and 
Eqs. (\ref{partialc1})-(\ref{partialc8}).
The radiative lifetimes of $4p_{1/2}$ and $4p_{3/2}$,  and $6p_{1/2}$ and 
$6p_{3/2}$ levels, are about 26 ns and 345 ns, respectively, and are included 
in the numerical calculations.

In  order to check the consistency of our  results an alternative formalism based on the
density matrix equations  was used  to calculate  the dynamics of the system.  
The amplitude equations approach has the advantages of dealing with fewer number of 
differential equations and obtaining more compact formulae for the ionization yield. 
On the other hand, the density matrix equations approach has the advantage of 
its capability to control the coherence through the off-diagonal density matrix elements  
(that are proportional to $\rho_{ij} = u_iu^*_j$, with $i \neq j$), and therefore
it is suitably used for mixed states (such as $4p_{1/2}(m_j=\pm 1/2)$ or $4p_{3/2}(m_j=\pm 1/2)$) 
when  at least two levels \textit{with arbitrary phase} are initially occupied.  
Details about the  density matrix approach are given in Appendix \ref{A}, and 
the numerical results are, of course, identical to the ones obtained by using 
the amplitude equations.    
In the following subsections we present numerical results for the 
K $4p_{1/2}-6p_{1/2}$ and $4p_{3/2}-6p_{3/2}$ systems.

\subsection{K $4p_{1/2}-6p_{1/2}$ system}

Figs. \ref{fig3}(a) and \ref{fig3}(b) show the variation of the total  ionization yield
and branching ratio as a function  of  two-photon detuning, $\delta$, at four different 
values of the polarization angle, $\theta_p=0^{\circ}$, $30^{\circ}$, $60^{\circ}$ 
and $90^{\circ}$. 
Note that the polarization angle of the dressing laser is fixed to $\theta_d=0^{\circ}$.
Pulse durations and peak laser intensities are chosen to be
$\tau_{p} = 1$ ns and $I_{p} = 1$ MW/cm$^{2}$, and $\tau_{d} = 10$ ns 
and $I_{d} = 100$ MW/cm$^{2}$, for the probe and dressing lasers, respectively.
 \noindent
Clearly, the profile of the LICS resonance and the branching ratio 
as a function of detuning for the $4p_{1/2}-6p_{1/2}$ system changes 
by varying the polarization angle.
The position of the LICS resonance is also altered by varying $\theta_p$, 
specifically the maximum of the LICS profile shifts toward larger values of the detuning 
as $\theta_p$ increases and its minimum vanishes completely when $\theta_p=90^{\circ}$. 
\noindent
At $\theta_p=90^{\circ}$ the value of the asymmetry parameter, which is connected 
to the resonance profile, $q=-6.59$, is much larger compared to the case when both 
lasers are linearly polarized in the same direction,  and  the asymmetry parameter 
takes the value, $q=-0.91$. That particular value of the asymmetry parameter for 
$\theta_p=90^{\circ}$ is due to the fact that the corresponding angular coefficients 
for the $s$ and $d$ ionization channels are equal, and the radial matrix elements 
have opposite signs.

Figs. \ref{fig4}(a) and \ref{fig4}(b) show the variation of the total  ionization yield 
and branching ratio as a function of detuning at three different dressing laser
intensities $I_{d} = 100$, $200$, and $500$ MW/cm$^{2}$, with the probe laser
intensity and the pulse durations fixed to be $I_{p}=100$ MW/cm$^{2}$,  
$\tau_{p} = 1$ ns,  and $\tau_{d} = 15$ ns.  
The polarization angle is $\theta_p=30^{\circ}$.
As we have already seen in Figs. \ref{fig3}(a)  and \ref{fig3}(b), 
the ionization yields and branching ratios vary significantly near resonance.  
The LICS structure is naturally broadened  as the dressing laser intensity is increased.
 
For the particular polarization geometry shown in Fig. \ref{fig1} the azimuthal
angle dependence, $\Phi$, of the  photoelectron signal, through the spherical
harmonics $Y_{lm}(\Phi, \Theta)$, does not vanish as it happens when the polarization 
axes of both lasers are  parallel to each other \cite{takgab}, and the cylindrical symmetry 
of PAD  is broken.  This  is due to the presence of the spherical harmonics 
with $m \neq 0$ in the differential ionization widths formulae 
Eqs. (\ref{pad12gamma13p})-(\ref{pad32gamma24m}).
 \noindent
Before studying  PAD for LICS it would be instructive to give an answer 
to the following question: 
What is the modification of PAD for one-photon ionization from the initial state 
$4p_{1/2}$ through the variation of the polarization angle of the probe laser 
without the dressing laser, i.e., $I_d=0$?
Since the initial  state, $4p_{1/2}$, is spherically symmetric (recall that both 
$m_j=\pm 1/2$ sublevels are equally populated), one could intuitively guess that 
the magnitude of PAD does not change, and PAD just aligns along the polarization 
axis of the probe laser. 
Under the condition of $I_d=0$, three-dimensional (3D) PAD is plotted in 
Figs. \ref{fig5}(a)-\ref{fig5}(c) as a function of photoelectron angles 
$\Theta$ and $\Phi$,  for the three different values of the polarization angle,  
$\theta_p=0^{\circ}$, $45^{\circ}$, and  $90^{\circ}$.
As  expected,  PAD changes its orientation along the polarization direction 
of the probe laser. The fourfold rotational symmetry ($\Phi\to \pi+\Phi$, and 
$\Phi \to -\Phi$, at $\Theta= 90^{\circ}$) that exists when both lasers are 
linearly polarized along the quantization axis, breaks into a twofold symmetry 
($\Phi\to \pi+\Phi$) when the polarization of the probe laser  varies \cite{bas,flo}.  

Now we consider the case of LICS, i.e., the dressing laser is turned on. 
3D PAD's of the K $4p_{1/2}-6p_{1/2}$ system at $\theta_p=60^{\circ}$ 
are shown in Fig. \ref{fig6}(a)-\ref{fig6}(c) for the three
representative detunings corresponding to the far-off resonance ($\delta = -4$ GHz),
maximum ($\delta = -0.62$ GHz), and minimum  ($\delta = 0.44$ GHz) of the branching
ratio (see Fig. \ref{fig3}(b)). 
 The view point of all  3D plots in this paper is from the $xy$-plane with the Cartesian 
 coordinates (2,2,0), if not otherwise stated.
At far-off resonance (Fig. \ref{fig6}(a)), the 3D PAD again tends to follow the
change of the polarization angle, $\theta_p=60^{\circ}$, with some small distortion
due to the dressing laser. The distortion, however, is almost invisible, since 
the interference effect through LICS is negligible at far-off resonance. 
In Figs. \ref{fig6}(b) and \ref{fig6}(c), we see that the 3D PAD's are significantly 
modified. Especially in Fig. \ref{fig6}(c), a maximum distortion is observed in 
PAD due to the strong destructive interference between the $\epsilon s$ and 
$\epsilon d$ partial waves. 

\subsection{K $4p_{3/2}-6p_{3/2}$ system}

We now turn to the case of the $4p_{3/2}-6p_{3/2}$ system.
This system is somehow different from  the $4p_{1/2}$-$6p_{1/2}$ system,  because
the initial state, $4p_{3/2}$, is not spherically symmetric: Only  
$m_j=\pm 1/2$ out of all possible $m_j=\pm 1/2,\; \pm 3/2 $ magnetic sublevels 
are equally occupied by the auxiliary laser, and for this reason a different 
behavior is expected. 

In Fig. \ref{fig7}(a) we plot the variation of the total  ionization yield
as a function of two-photon detuning at four different values of the polarization
angle, $\theta_p=0^{\circ}$, $30^{\circ}$, $60^{\circ}$, and $90^{\circ}$.   
The LICS structure in Fig. \ref{fig7}(a) is  not quite similar to that plotted 
in Fig. \ref{fig3}(a), because,  although $q_{13}^{\epsilon s}$ and 
$q_{13}^{\epsilon d}$ are the same for both systems,  $q_{13}$ itself is different. 
More interestingly, the variation of the branching ratios shown in Fig. \ref{fig7}(b)
is substantially larger than that shown in Fig. \ref{fig3}(b) as $\theta_p$ increases. 
For $\theta_p=90^{\circ}$ the ionization yield into the $\epsilon d$ continuum 
at the detuning close to $\delta= -0.9 $ GHz is almost 40 times enhanced compared 
to that into the $\epsilon s$ continuum. 
This suggests that an appropriate choice of the probe polarization angle 
and the two-photon detuning leads to the control of ionization into different channels.
Recent experiments \cite{aloise} performed for ionization from an excited state 
of Xe with linearly and circularly polarized lasers have demonstrated that 
the ionization products into different continua can be separated by 
varying the polarization of lasers. 
The variation of the total ionization yields and the branching ratios 
as a function of detuning $\delta$ at three different dressing laser intensities, 
 $I_{d} = 100$, $200$, and $500$ MW/cm$^{2}$, is presented in Figs. \ref{fig8}(a) 
 and  \ref{fig8}(b) with the rest of the parameters being the same as those in Fig. \ref{fig4}.
 
In Figs. \ref{fig9}(a)-\ref{fig9}(c) we plot the 3D PAD for one-photon 
ionization from the $4p_{3/2}$ state by the probe laser
at $\theta_p=0^{\circ}$, $45^{\circ}$, and $90^{\circ}$, without the dressing 
laser, i.e., $I_d=0$.
Compared to  the $4p_{1/2}-6p_{1/2}$ system (see Figs. \ref{fig5}(a)-\ref{fig5}(c)), 
PAD's drastically change the shape with the
detuning when the probe polarization angle is varied. 

Now we return to the case for LICS by turning on the dressing laser, and present the 
3D PAD's for the $4p_{3/2}-6p_{3/2}$ system in Fig. \ref{fig10}(a)-\ref{fig10}(c), 
at $\theta_p=60^{\circ}$, for three representative detunings corresponding 
to the far-off resonance ($\delta = -4$ GHz),  
maximum ($\delta = -0.89$ GHz), and minimum ($\delta = 0.42$ GHz) of the 
branching ratio (see Fig. \ref{fig7}(b)). 
The modification of the 3D PAD's, presented in Figs. \ref{fig9} and \ref{fig10}, 
is more than we expect: The variation of the sidelobes of 3D PAD's at different 
polarization angles, which are absent for the $4p_{1/2}-6p_{1/2}$ system, is striking. 
The sidelobes are due to the ionization into the $\epsilon d_{5/2}$ continuum 
(that is inaccessible through one-photon ionization from the $4p_{1/2}$ state), 
and are present even if $\theta_p=0^{\circ}$.

As we have already noticed, not only the photoelectron angular distribution but also 
the angle-integrated ionization yield is affected by the relative polarization 
angle between the probe and dressing lasers. This effect is called 
{\it linear dichroism} and is very attractive from the experimental point of view, 
since it is much easier to measure the total ionization yield than the PAD. 
Linear dichroism can be experimentally used to determine the ratio of the 
dipole matrix elements into the different continua \cite{kab}, or the 
relative phase shift between the partial waves of the continua.   
The normalized linear dichroism is defined as \cite{che}, 
\bqn
LD=\frac{R(\theta_p=90^{\circ}) -R(\theta_p=0^{\circ}) }
	{R(\theta_p=90^{\circ}) +R(\theta_p=0^{\circ})},
\eqn 
where $R(\theta_p=0^{\circ}) $ and $R(\theta_p=90^{\circ}) $ represent the 
total ionization yield, or equivalently the angle-integrated photoelectron signal 
when the polarization axis of the probe laser is parallel and perpendicular with 
respect to that of the dressing laser, respectively.
In Figs. \ref{fig11}(a) and \ref{fig11}(b) we plot the linear dichroism, $LD$, 
as a function of two-photon detuning for the K $4p_{1/2}-6p_{1/2}$ and 
$4p_{3/2}-6p_{3/2}$ systems, respectively.
The magnitude of the linear dichroism changes drastically for both systems 
around the LICS resonance, and it shows a large maximum for the 
two-photon detunings around the deep LICS minimum at $\theta_p=0^{\circ}$.

\section{SUMMARY} 

In this paper we have theoretically investigated the effects of the relative 
polarization angle between the probe and dressing lasers on the total 
(angle-integrated) ionization yield, branching ratio, and PAD through LICS 
for the K $4p_{1/2}-6p_{1/2}$ and $4p_{3/2}-6p_{3/2}$ systems in a 
particular geometry with both probe and dressing lasers being linearly polarized. 
Amplitude equations and alternatively density matrix equations formalisms 
have been used to study the dynamics of the ionization process.   
We have shown that the ionization yield and the branching ratio are 
strongly dependent on the relative polarization angle between the lasers. 
Moreover we have found that ionization into the different continua, 
branching ratios, and PAD are significantly altered by the change of the 
polarization angle. 
Our findings suggest that the relative polarization angle can be another 
doorknob to control the ionization dynamics through LICS. 
We have also calculated linear dichroism for the angle-integrated 
ionization yield, which turned out to be quite large at the two-photon detunings 
close to the LICS minimum.

\acknowledgements{GB acknowledges financial support from Japan Society for the 
Promotion of Science (JSPS).   The work by TN was supported by the Grant-in-Aid
for scientific research from the Ministry of Education and Science of Japan.}

\appendix
\section{Time-dependent density matrix equations}\label{A}

Based on the density matrix approach \cite{dixlamb}, we study the temporal 
evolution of the K atom in the laser field given by Eq. (\ref{field}). 
Briefly we solve the following set of time-dependent differential equations 
for the slowly varying density matrix $\sigma$: 
\bqna
&&\dot{\sigma}_{ii} =   - \widetilde{\Gamma}_i  {\sigma}_{ii} 
	-2 {\rm Im}\left[  \sum_{j=3}^{4} 
	\Omega_{ji} \left( 1 + \frac{i}{q_{ji}} \right) {\sigma}_{ij}  \right]	
             \;,
\label{dm1} \\
&&\dot{\sigma}_{jj} =   - \widetilde{\Gamma}_j  {\sigma}_{jj} 
	+2 {\rm Im}\left[  \sum_{i=1}^{2} 
	\Omega_{ji} \left( 1 - \frac{i}{q_{ji}} \right) {\sigma}_{ij}  \right] 
	 \;,  
\label{dm2} \\
&&\dot{\sigma}_{ij} =  
\left[
i\delta_{ij}-\frac{1}{2}(\widetilde{\Gamma}_i+ \widetilde{\Gamma}_j )
\right] {\sigma}_{ij} 	
	+ i  \sum_{i'=1}^{2} 
	\Omega_{i'j} \left( 1 - \frac{i}{q_{i'j}} \right) {\sigma}_{ii'}
	- i  \sum_{j'=3}^{4} 
	\Omega_{ij'} \left( 1 - \frac{i}{q_{ij'}} \right) {\sigma}_{j'j}     
	\;,
\label{dm3}\\
&&\dot{\sigma}_{ii'} =  
-\frac{1}{2}(\widetilde{\Gamma}_i+ \widetilde{\Gamma}_{i'} ) {\sigma}_{ii'} 	
	+ i  \sum_{j=3}^{4} 
	\Omega_{ji'} \left( 1 + \frac{i}{q_{ji'}} \right) {\sigma}_{ij}
	- i  \sum_{j=3}^{4} 
	\Omega_{i'j} \left( 1 - \frac{i}{q_{i'j}} \right) {\sigma}_{ji'}   
 \;,  
\label{dm4}\\
&&\dot{\sigma}_{jj'} =  
-\frac{1}{2}(\widetilde{\Gamma}_j+ \widetilde{\Gamma}_{j'} ) {\sigma}_{jj'} 	
	+ i  \sum_{i=1}^{2} 
	\Omega_{ij'} \left( 1 + \frac{i}{q_{ij'}} \right) {\sigma}_{ji}
	- i  \sum_{i=1}^{2} 
	\Omega_{ji} \left( 1 - \frac{i}{q_{ji}} \right) {\sigma}_{ij'} 
	\;,	  
\label{dm5}
\eqna
\noindent
where the indices take the following values $i,i'=1,2$  and $j,j'=3,4$, 
with $i\neq i'$ and $j\neq j'$.
All the density matrix elements for the continuum have been adiabatically 
eliminated from the Eqs. (\ref{dm1})-{(\ref{dm5}).  
Note that we have used the rotating wave approximation and the slowly varying 
density matrix elements to derive the above equations: 
${\sigma}_{ii}= \rho_{ii}$, ($i=1,4$),
${\sigma}_{ij}= \rho_{ij}e^{-i\delta_{static,\;ij} t}$, ($i={1,2}$, and $j={3,4}$), 
${\sigma}_{ic}= \rho_{ic}e^{-i\omega_{\alpha} t}$, ($\alpha=p$ or $d$, and  $i=1, 4$),
where  $\rho_{ij}(t)= u_{i}(t) u_{j}^*(t)$ are the density matrix elements.
$\delta_{ij}$ is the two-photon detuning defined by $\delta_{ij}= \delta_{static,\;ij} +
\delta_{stark,\;ij}$, where the static detuning is defined by
$\delta_{static,\;ij} = ( E_{i} + \hbar \omega_{p} ) - ( E_{j} + \hbar \omega_{d} )$, 
and $\delta_{stark,\;ij}$ is the total dynamic ac Stark shift defined by
$\delta_{stark,\;ij}=(S_{i}^{(p)} + S_{i}^{(d)}) - (S_{j}^{(p)} + S_{j}^{(d)})$. 
Now the above set of density matrix equations is solved with the following 
initial conditions:
$\sigma_{ii}(t=-\infty) =1/2$, and $\sigma_{jj}(t=-\infty) =\sigma_{ij}(t=-\infty)=0$, 
for $i=1,2$ and   $j=3,4$.
\noindent
The total (angle-integrated) ionization yield is derived as, 
\bqn
R(t) =\int_{-\infty}^{t}dt'\left\{ 
 \sum_{i=1}^{2}\Gamma_{i}^{(p)}  \sigma_{ii} + 
 \sum_{j=3}^{4}(\Gamma_{j}^{(d)} + \Gamma_{j}^{(p)} )  \sigma_{jj}
 +4 \sum_{i=1}^{2} \sum_{j=3}^{4} \;{\rm Im}
	{\left[\Omega_{ji}\left(1+\frac{i}{q_{ji}}\right)\right]}
 		{\rm Re}(\sigma_{ij})
 \right\}. 
 \label{totalyielddm}
\eqn
\noindent
The probability that a photoelectron is ejected into a solid angle,  
$\Omega_{\textbf{k}}$, is given by the following formula:
\noindent
\bqna
\frac{dR (\Theta,\Phi)}{dt d\Omega_{\textbf{k}}} &=& 
0.589 \;\pi \sum_{m_{s'}=\pm 1/2}
 \left\{
   \sum_{i=1}^{2}\;
       {\Gamma_{i}^{(p,m_{s'})}(\Theta,\Phi)} \;\sigma_{ii}                      
\right. \nonumber\\ &+& \left.    
   \sum_{j=3}^{4}
    \left[\Gamma_{j}^{(d,m_{s'})}(\Theta,\Phi)
    + \Gamma_{j}^{(p,m_{s'})}(\Theta,\Phi)\right] \;\sigma_{jj} 
 \right. \nonumber\\ &+& \left.      
      2 {\rm Re}
    \left[ \sum_{i=1}^{2} \sum_{j=3}^{4}
    \sqrt{ \Gamma_{i}^{(p,m_{s'})} (\Theta,\Phi) }
   \sqrt{   \left( \Gamma_{j}^{(p,m_{s'})}(\Theta,\Phi) \right)^*}
   \sigma_{ij}  \right]  
    \right. \nonumber\\ &+& \left.    
   2 {\rm Re}
       \left[ \sqrt{ \Gamma_{1}^{(p,m_{s'})} (\Theta,\Phi) }
       \sqrt{  \left( \Gamma_{2}^{(p,m_{s'})}(\Theta,\Phi) \right)^*}
   \sigma_{12}\right]
    \right. \nonumber\\ &+& \left.    
    2 {\rm Re}
       \left[ \sqrt{ \Gamma_{3}^{(d,m_{s'})} (\Theta,\Phi) }
      \sqrt{   \left( \Gamma_{4}^{(d,m_{s'})}(\Theta,\Phi) \right)^*}
   \sigma_{34} \right]
  \right. \nonumber\\ &+& \left.    
    2 {\rm Re}
       \left[ \sqrt{ \Gamma_{3}^{(p,m_{s'})} (\Theta,\Phi) }
      \sqrt{  \left( \Gamma_{4}^{(p,m_{s'})}(\Theta,\Phi) \right)^*}
   \sigma_{34} \right]  
\right\}
\;,
\label{paddm}
\eqna
which can be shown to be equivalent with Eq. (\ref{pad_total}). 
The numerical results  for both K $4p_{1/2}-6p_{1/2}$ and $4p_{3/2}-6p_{3/2}$ systems 
obtained in the density matrix and amplitude equations formalisms are, 
of course, identical.  

\section{Ionization widths} \label{B}

The partial ionization width  from state $|j\rangle$ to the continuum $|c\rangle$ 
produced by laser $\alpha$, is defined as:
\bqn
\Gamma_{jc}^{(\alpha,m_{s'})}(\Theta,\Phi) =  
	2 \pi |D_{jc}^{(\alpha,m_{s'})}(\Theta,\Phi)|^{2} , 
\eqn
where $D_{jc}^{(\alpha,m_{s'})}= -{\rm E}_{\alpha}(t) 
\sum_{q=\pm 1,0} \langle c|{r}_q {e}_{q}^{(\alpha)}|j\rangle \equiv {\rm
E}_{\alpha}(t) \mu_{jc}^{(\alpha,m_{s'})} $
represents the one-photon dipole matrix element between states $|j\rangle$ and 
$|c\rangle$, expressed in the length gauge, and calculated at energy
$E_c= E_j+\omega_{\alpha}$. 
The total ionization width integrated over the solid angle, $\Omega_\textbf{k}$, 
defined by the polar angles ($\Theta$, $\Phi$), of the ejected photoelectron 
is given by, 
\bqn
\Gamma_{j}^{(\alpha)}= \sum_{m_{s'}=\pm 1/2}
\sum_{c } \Gamma_{jc}^{(\alpha,m_{s'})} .
\eqn
The summation over $c$ implies that the summation is taken over the all 
allowed continuum states.
 
\section{Ac Stark shifts}\label{C}

The dynamic ac Stark shift of the energy of state $|j\rangle$ due to 
both bound and free states $|k\rangle$, caused by laser  $\alpha$, 
 is, 
\bqn
S_{j}^{(\alpha)} = 
 \sum_{m_{s'}=\pm 1/2} \sum_{k} 
\left[
\frac{ |D_{jk}^{(\alpha,m_{s'})}|^{2}}
{E_j+\omega_{\alpha}-E_k +i\varepsilon}
+
\frac{ |D_{jk}^{(\alpha,m_{s'})}|^{2}}
{E_j-\omega_{\alpha}-E_k +i\varepsilon}
\right].
\eqn
The sum here contains both summation over the bound  and integration over 
the continuum states, and $\varepsilon$ is an infinitely  small number.

\section{Two-photon Rabi frequency }\label{D}

The total  two-photon Rabi frequency $\Omega_{ij}$ between the state
$|i\rangle$ and $|j\rangle$ is given by, 
\bqn
\Omega_{ij} \left( 1 - \frac{i}{q_{ij}} \right)=  \sum_{c}
 \Omega_{ij}^{c} \left( 1 - \frac{i}{q_{ij}^{c}} \right),
\eqn
where the partial two-photon Rabi frequency between  state
$|i\rangle$ and $|j\rangle$ coupled through the continuum $|c\rangle$ with the energy
$E_{1}+\omega_p \simeq  E_{3}+\omega_d$ is defined as, 
\bqn
\Omega_{ij}^{c} \left( 1 - \frac{i}{q_{ij}^c} \right) = 
\sum_{m_{s'}=\pm 1/2}\int dE_c 
\frac{ D_{ic}^{(p,m_{s'})}D_{cj}^{(d,m_{s'})}}
{E_1+\omega_p-E_{c}  +i\varepsilon}.
\eqn
The imaginary part of the partial Rabi frequency  is connected to the partial 
asymmetry parameter $q_{ij}^{c}$, and is given by, 
\bqn
\left.
\frac{ \Omega_{ij}^{c}}{q_{ij}^{c}}= \sum_{m_{s'}=\pm 1/2}  \pi  
D_{ic}^{(p,m_{s'})}D_{cj}^{(d,m_{s'})}(\Theta,\Phi) \right\vert_
{E_{c}=E_1+\omega_ p}.
\eqn

\newpage

\begin{center}
\begin{tabular}{lp{5in}}
Table I. & Atomic parameters for the K $4P_{1/2}$-$6p_{1/2}$ system. 
$\Omega$ is measured in rad/s, $\Gamma$ in s$^{-1}$, $S$ in rad/s, 
and $I_{d}$ in W/cm$^{2}$.\\
\\
\end{tabular}

\begin{tabular}{c c c c c c c c c c c c c c c}
\hline\hline
  \multicolumn{1}{c}{$\Omega_{13}$} 
&&&&\multicolumn{1}{c}{$-8.12 \sqrt{I_{p} I_{d}} \cos \theta_p$} 
&&&&&&\multicolumn{1}{c}{$q_{13}$} 
&&&&\multicolumn{1}{c}{$-0.91$} \\
  \multicolumn{1}{c}{$\Omega_{13}^{\epsilon s}$} 
&&&&\multicolumn{1}{c}{$3.47 \sqrt{I_{p} I_{d}}\cos \theta_p$} 
&&&&&&\multicolumn{1}{c}{$q_{13}^{\epsilon s}$} 
&&&&\multicolumn{1}{c}{$1.71$} \\
  \multicolumn{1}{c}{$\Omega_{13}^{\epsilon d}$}   
&&&&\multicolumn{1}{c}{$-11.58 \sqrt{I_{p} I_{d}} \cos \theta_p$} 
&&&&&&\multicolumn{1}{c}{$q_{13}^{\epsilon d}$}       
&&&&\multicolumn{1}{c}{$-1.69$} \\
\hline
 \multicolumn{1}{c}{$\Omega_{14}$} 
&&&&\multicolumn{1}{c}{$-9.26 \sqrt{I_{p} I_{d}} \sin \theta_p$} 
&&&&&&\multicolumn{1}{c}{$q_{14}$} 
&&&&\multicolumn{1}{c}{$-6.59$} \\
  \multicolumn{1}{c}{$\Omega_{14}^{\epsilon s}$} 
&&&&\multicolumn{1}{c}{$3.47 \sqrt{I_{p} I_{d}}\sin \theta_p$} 
&&&&&&\multicolumn{1}{c}{$q_{14}^{\epsilon s}$} 
&&&&\multicolumn{1}{c}{$1.71$} \\
  \multicolumn{1}{c}{$\Omega_{14}^{\epsilon d}$}   
&&&&\multicolumn{1}{c}{$-5.79 \sqrt{I_{p} I_{d}} \sin \theta_p$} 
&&&&&&\multicolumn{1}{c}{$q_{14}^{\epsilon d}$}       
&&&&\multicolumn{1}{c}{$-1.69$} \\
\hline
  \multicolumn{1}{c}{$\Gamma_{1}^{(p)}$} 
&&&&\multicolumn{1}{c}{$11.59 I_{p}$}
&&&&&&\multicolumn{1}{c}{$S_{1}^{(p)}$} 
&&&&\multicolumn{1}{c}{$14.1 I_{p}$} \\
  \multicolumn{1}{c}{$\Gamma_{3}^{(d)}$} 
&&&&\multicolumn{1}{c}{$28.04 I_{d}$} 
&&&&&&\multicolumn{1}{c}{$S_{1}^{(d)}$} 
&&&&\multicolumn{1}{c}{$947.5 I_{d}$} \\
  \multicolumn{1}{c}{$\Gamma_{3}^{(p)}$} 
&&&&\multicolumn{1}{c}{$3.66I_{p}$}
&&&&&&\multicolumn{1}{c}{$S_{3}^{(p)}$} 
&&&&\multicolumn{1}{c}{$21.04 I_{p}$}\\
  \multicolumn{1}{c}{} 
&&&&\multicolumn{1}{c}{} 
&&&&&&\multicolumn{1}{c}{$S_{3}^{(d)}$} 
&&&&\multicolumn{1}{c}{$86.9 I_{d}$} \\
\hline\hline
\end{tabular}
\end{center}

\begin{center}
\begin{tabular}{lp{5in}}
Table II. & Atomic parameters for the K $4P_{3/2}$-$6p_{3/2}$ system. 
$\Omega$ is measured in rad/s, $\Gamma$ in s$^{-1}$, $S$ in rad/s, 
and $I_{d}$ in W/cm$^{2}$.\\
\\
\end{tabular}

\begin{tabular}{c c c c c c c c c c c c c c c}
\hline\hline
  \multicolumn{1}{c}{$\Omega_{13}$} 
&&&&\multicolumn{1}{c}{$-5.80 \sqrt{I_{p} I_{d}}\cos \theta_p$} 
&&&&&&\multicolumn{1}{c}{$q_{13}$} 
&&&&\multicolumn{1}{c}{$-0.5$} \\
  \multicolumn{1}{c}{$\Omega_{13}^{\epsilon s}$} 
&&&&\multicolumn{1}{c}{$6.94 \sqrt{I_{p} I_{d}}\cos \theta_p$} 
&&&&&&\multicolumn{1}{c}{$q_{13}^{\epsilon s}$} 
&&&&\multicolumn{1}{c}{$1.71$} \\
  \multicolumn{1}{c}{$\Omega_{13}^{\epsilon d}$}   
&&&&\multicolumn{1}{c}{$-12.74 \sqrt{I_{p} I_{d}}\cos \theta_p$} 
&&&&&&\multicolumn{1}{c}{$q_{13}^{\epsilon d}$}       
&&&&\multicolumn{1}{c}{$-1.69$} \\
\hline
  \multicolumn{1}{c}{$\Omega_{14}$} 
&&&&\multicolumn{1}{c}{$-9.26 \sqrt{I_{p} I_{d}}\sin \theta_p$} 
&&&&&&\multicolumn{1}{c}{$q_{14}$} 
&&&&\multicolumn{1}{c}{$-6.59$} \\
  \multicolumn{1}{c}{$\Omega_{14}^{\epsilon s}$} 
&&&&\multicolumn{1}{c}{$3.47 \sqrt{I_{p} I_{d}}\sin \theta_p$} 
&&&&&&\multicolumn{1}{c}{$q_{14}^{\epsilon s}$} 
&&&&\multicolumn{1}{c}{$1.71$} \\
  \multicolumn{1}{c}{$\Omega_{14}^{\epsilon d}$}   
&&&&\multicolumn{1}{c}{$-5.79\sqrt{I_{p} I_{d}}\sin \theta_p$} 
&&&&&&\multicolumn{1}{c}{$q_{14}^{\epsilon d}$}       
&&&&\multicolumn{1}{c}{$-1.69$} \\
\hline
  \multicolumn{1}{c}{$\Gamma_{1}^{(p)}$} 
&&&&\multicolumn{1}{c}{$(14.46\cos^2 \theta_p +10.15 \sin^2 \theta_p ) I_{p}$}
&&&&&&\multicolumn{1}{c}{$S_{1}^{(p)}$} 
&&&&\multicolumn{1}{c}{$(12.3 \cos^2 \theta_p +15.03 \sin^2 \theta_p ) I_{p}$ } \\
  \multicolumn{1}{c}{$\Gamma_{3}^{(d)}$} 
&&&&\multicolumn{1}{c}{$38.57 I_{d}$} 
&&&&&&\multicolumn{1}{c}{$S_{1}^{(d)}$} 
&&&&\multicolumn{1}{c}{$1231.8 I_{d}$} \\
  \multicolumn{1}{c}{$\Gamma_{3}^{(p)}$} 
&&&&\multicolumn{1}{c}{$(4.32 \cos^2 \theta_p+3.33\sin^2 \theta_p)I_{p}$} 
&&&&&&\multicolumn{1}{c}{$S_{3}^{(p)}$} 
&&&&\multicolumn{1}{c}{$(20.8 \cos^2 \theta_p +21.15 \sin^2 \theta_p ) I_{p} $} \\
  \multicolumn{1}{c}{} 
&&&&\multicolumn{1}{c}{} 
&&&&&&\multicolumn{1}{c}{$S_{3}^{(d)}$} 
&&&&\multicolumn{1}{c}{$97.0 I_{d}$} \\
\hline\hline
\end{tabular}
\end{center}

\newpage

\begin{figure}
\centering
\includegraphics[width=3.in,angle=0]{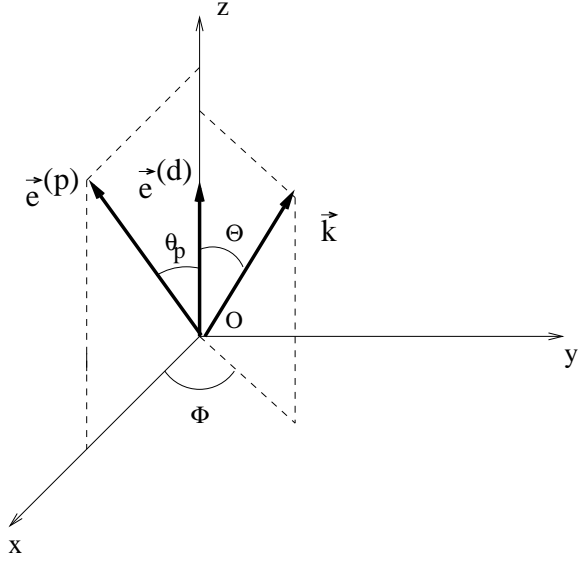}
\caption{
Quantization axis and the polarization vectors, $\textbf{e}^{(p)}$ and
$\textbf{e}^{(d)}$, for the probe and dressing lasers defined for this work.
The polarization vector $\textbf{e}^{(p)}$ lies in the $xz$-plane, 
and the quantization axis is taken along the $z$ axis.}
\label{fig1}
\end{figure}

\begin{figure}
\centering
\includegraphics[width=3.in,angle=0]{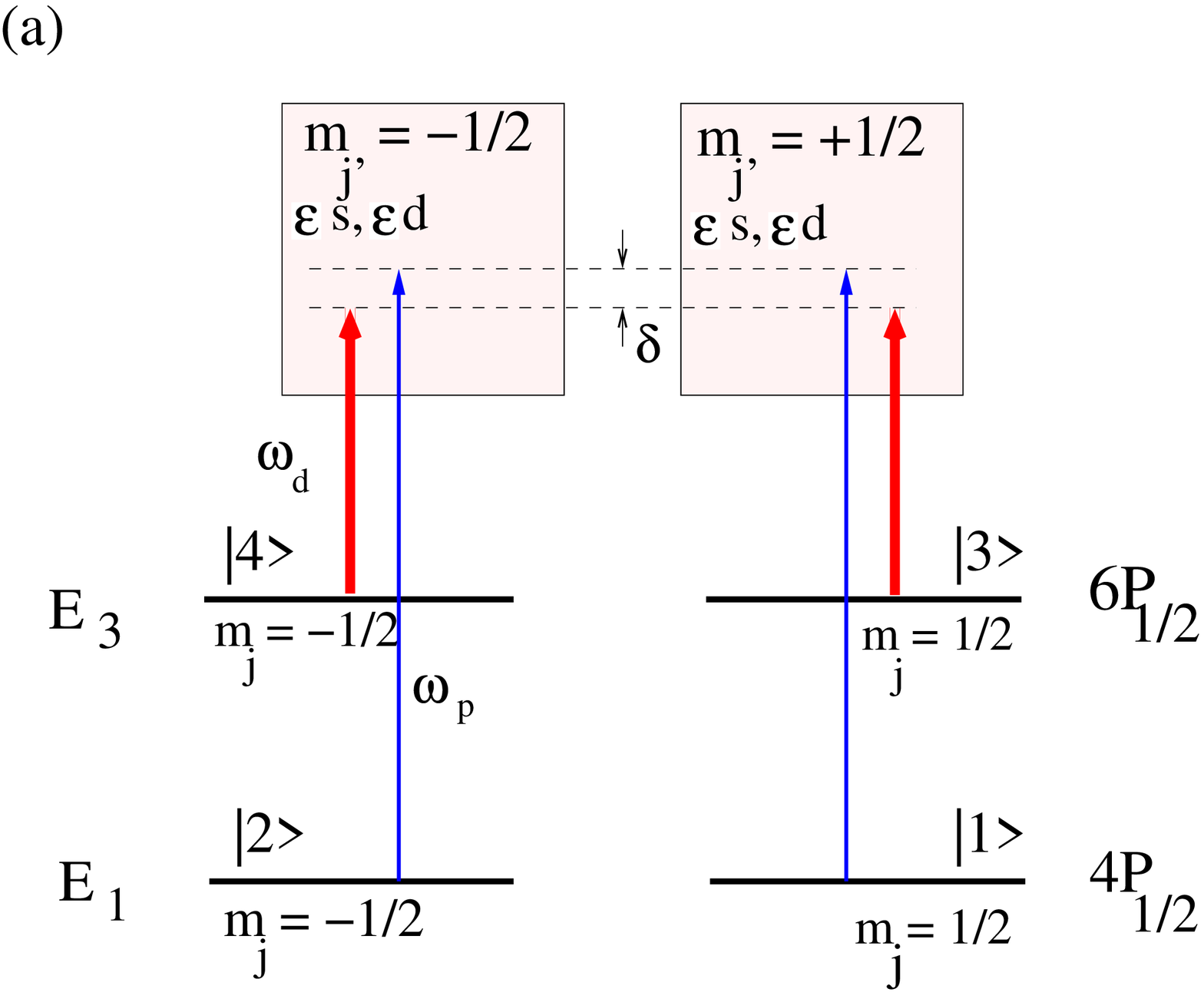}\\ 
\vspace{.5cm}
\includegraphics[width=3.in,angle=0]{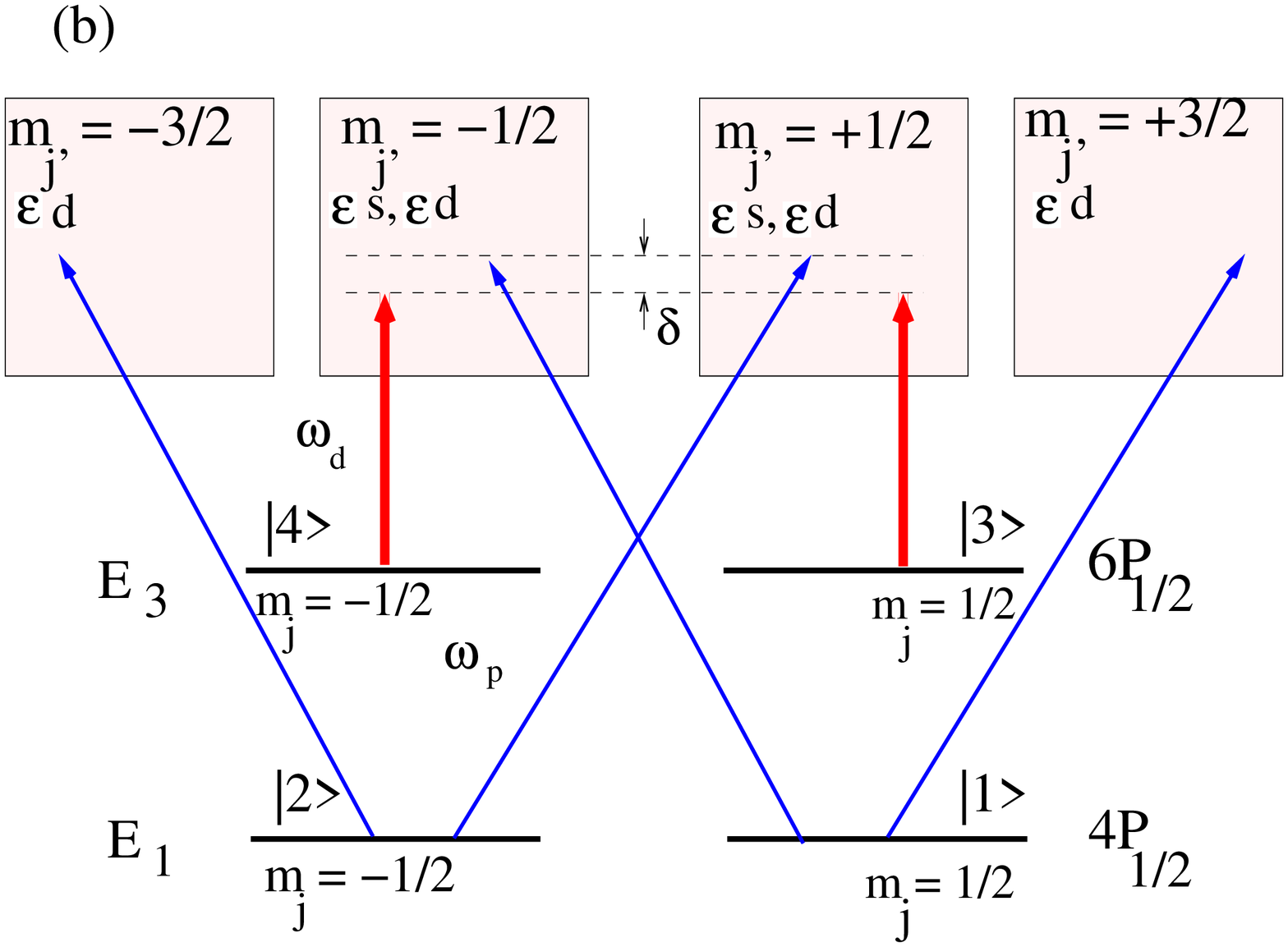}\\
\vspace{.5cm}
\includegraphics[width=3.in,angle=0]{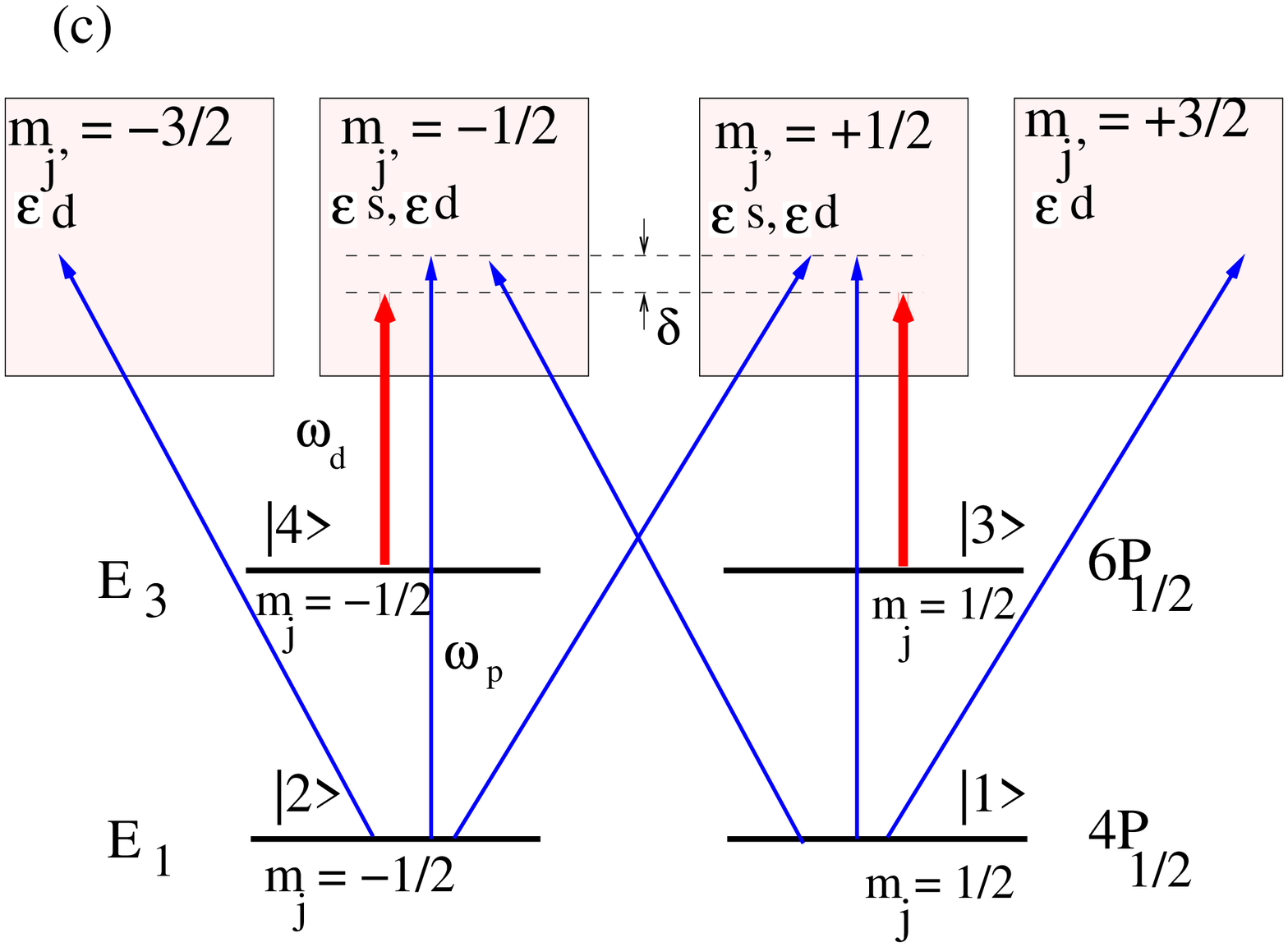}
\caption{Color online. Level scheme considered in this paper for the K $4p_{1/2}-6p_{1/2}$ 
system. Depending on the polarization angle of the probe laser, $\theta_{p}$,
different transition paths have to be considered. 
(a) $\theta_p=0^{\circ}$, (b) $\theta_p= 90^{\circ}$, and 
(c) $ 0^{\circ}< \theta_p< 90^{\circ}$.  Similar level scheme can be
drawn for the K $4p_{3/2}-6p_{3/2}$ system. 
}
\label{fig2}
\end{figure}

\begin{figure}
\centering
\includegraphics[width=6.0in,angle=0]{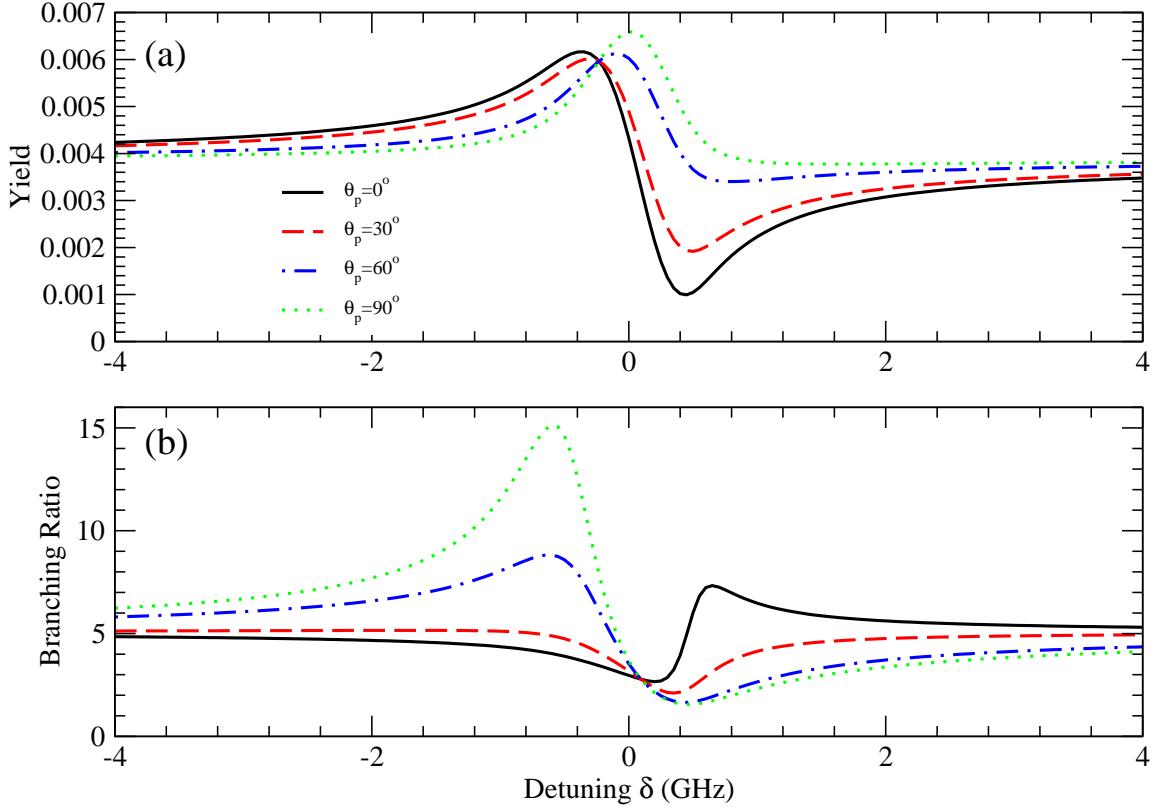}
\caption{Color online. 
(a) Total ionization yield and (b) the branching ratio between the partial 
ionization yields into each $\epsilon s$ and $\epsilon d$ continuum
for the K $4p_{1/2}-6p_{1/2}$ system as a function of two-photon detuning 
$\delta$.  Pulse durations and peak laser intensities are chosen to be
$\tau_{p} = 1$ ns and $I_{p} = 1 $ MW/cm$^{2}$, and $\tau_{d} = 10$ ns 
and $I_{d} = 100$ MW/cm$^{2}$, for the probe and dressing lasers, respectively.
The polarization angle takes the values of 
$\theta_p=0^{\circ}, 30^{\circ}, 60^{\circ}$ and $90^{\circ}$.
}
\label{fig3}
\end{figure}

\begin{figure}
\centering
\includegraphics[width=6.0in,angle=0]{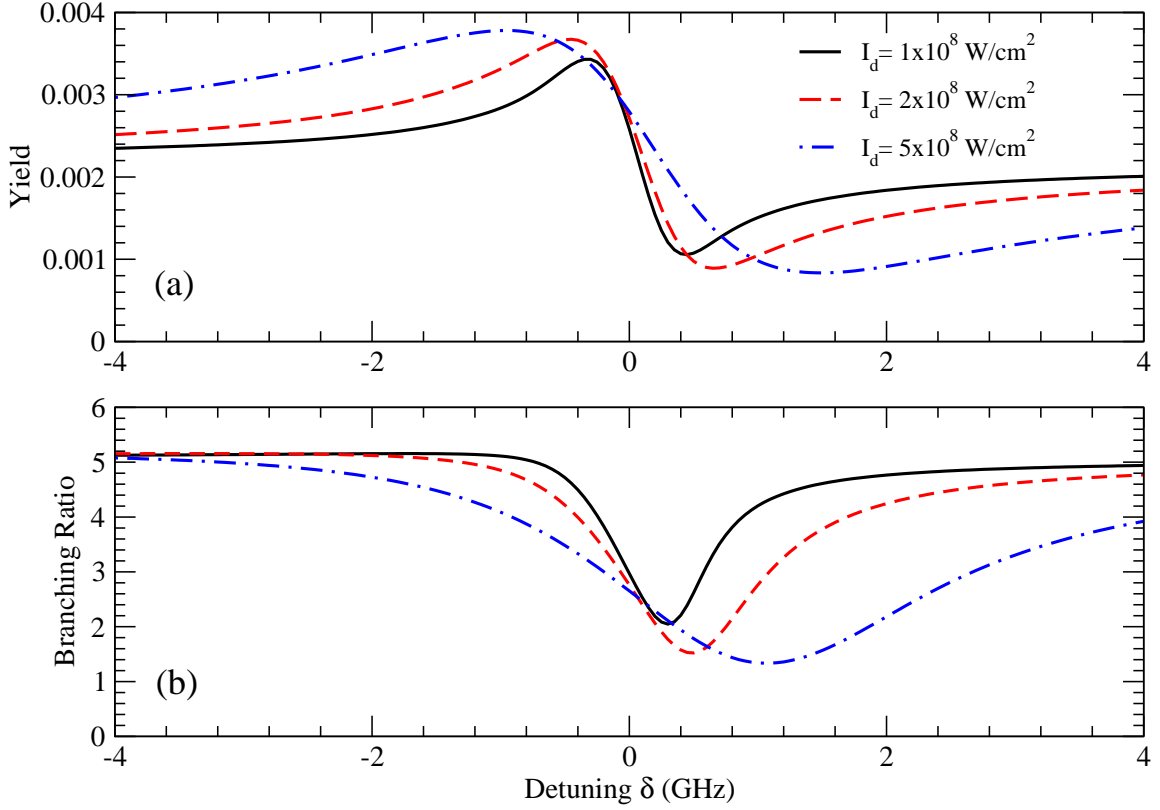}
\caption{Color online. 
(a) Total ionization yield and (b) the branching ratio between the partial 
ionization yields into each $\epsilon s$ and  $\epsilon d$ continuum
for the K $4p_{1/2}-6p_{1/2}$ system as a function of two-photon detuning 
$\delta$ for the three different dressing laser intensities, $I_{d}= 100$, $200$, 
and $500$ MW/cm$^{2}$. The intensity of the probe laser is   $I_{d}= 1$ MW/cm$^{2}$.
Pulse durations are chosen to be $\tau_{p} = 1$ ns and $\tau_{d} = 15$ ns, 
for the probe and dressing lasers, respectively. 
The polarization angle is $\theta_p=30^{\circ}$. 
}
\label{fig4}
\end{figure}

\newpage\begin{figure}
\centering
\includegraphics[width=2.0in,angle=0]{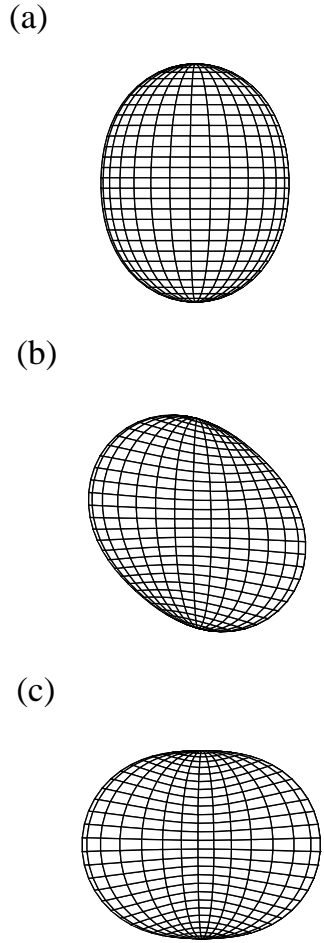}
\caption{
Three-dimensional photoelectron angular distribution due to one-photon ionization 
from the K $4p_{1/2}$ state by the probe laser field only, at three different 
polarization angles $\theta_p= 0^{\circ}$, $ 45^{\circ}$, and $ 90^{\circ}$.
Pulse duration and peak intensity are $\tau_{p}=1$ ns and 
$I_{p}= 1$ MW/cm$^{2}$ for the probe laser. 
The view point is from the positive $y$-axis.}
\label{fig5}
\end{figure}

\begin{figure}
\centering
\includegraphics[width=2.0in,angle=0]{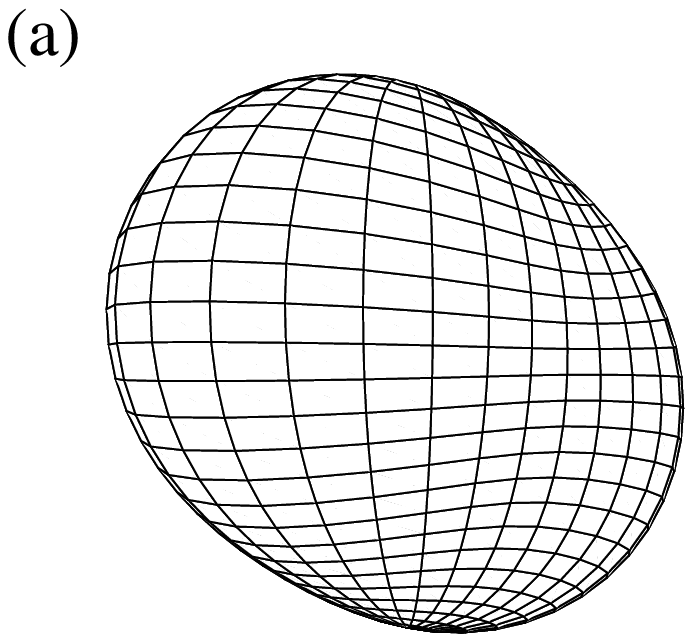}\\
\includegraphics[width=2.0in,angle=0]{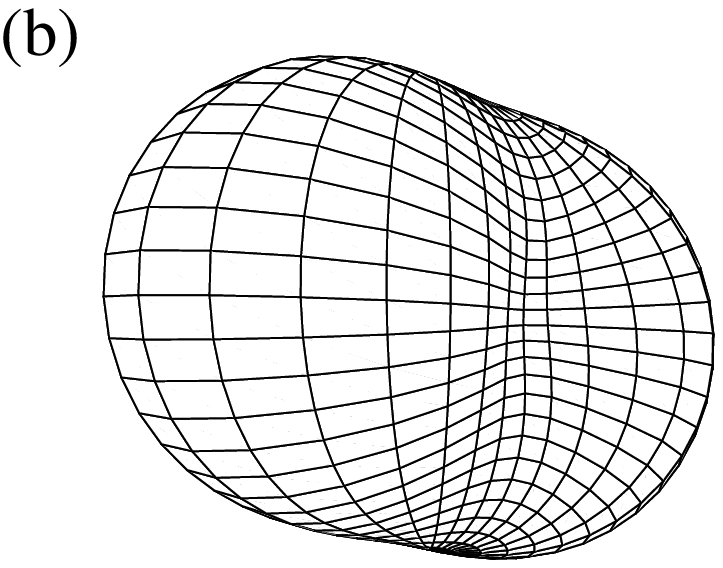}\\
\includegraphics[width=2.0in,angle=0]{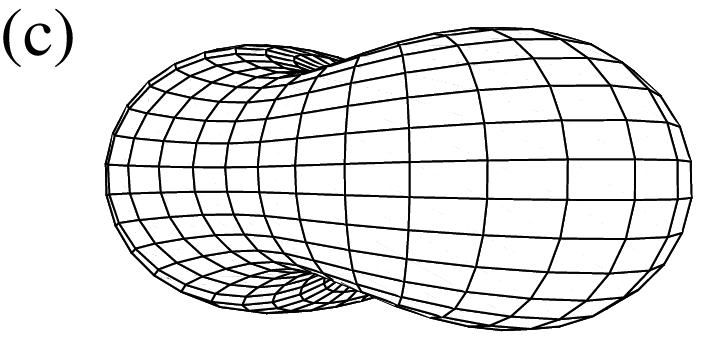}\\
\caption{Three-dimensional photoelectron angular distribution for the 
K $4p_{1/2}-6p_{1/2}$ system at three different two-photon detunings 
$\delta= -4$, $-0.62$, and $0.44$ GHz.  
Pulse durations and peak intensities are $\tau_{p}=1$ ns and 
$I_{p}= 1$ MW/cm$^{2}$ for the probe laser, and $\tau_{d}=10$ ns 
and $I_{p}= 100$ MW/cm$^{2}$ for the dressing laser. 
The polarization angle is $\theta_p= 60^{\circ}$.
}
\label{fig6}
\end{figure}

\newpage

\begin{figure}
\centering
\includegraphics[width=6.0in,angle=0]{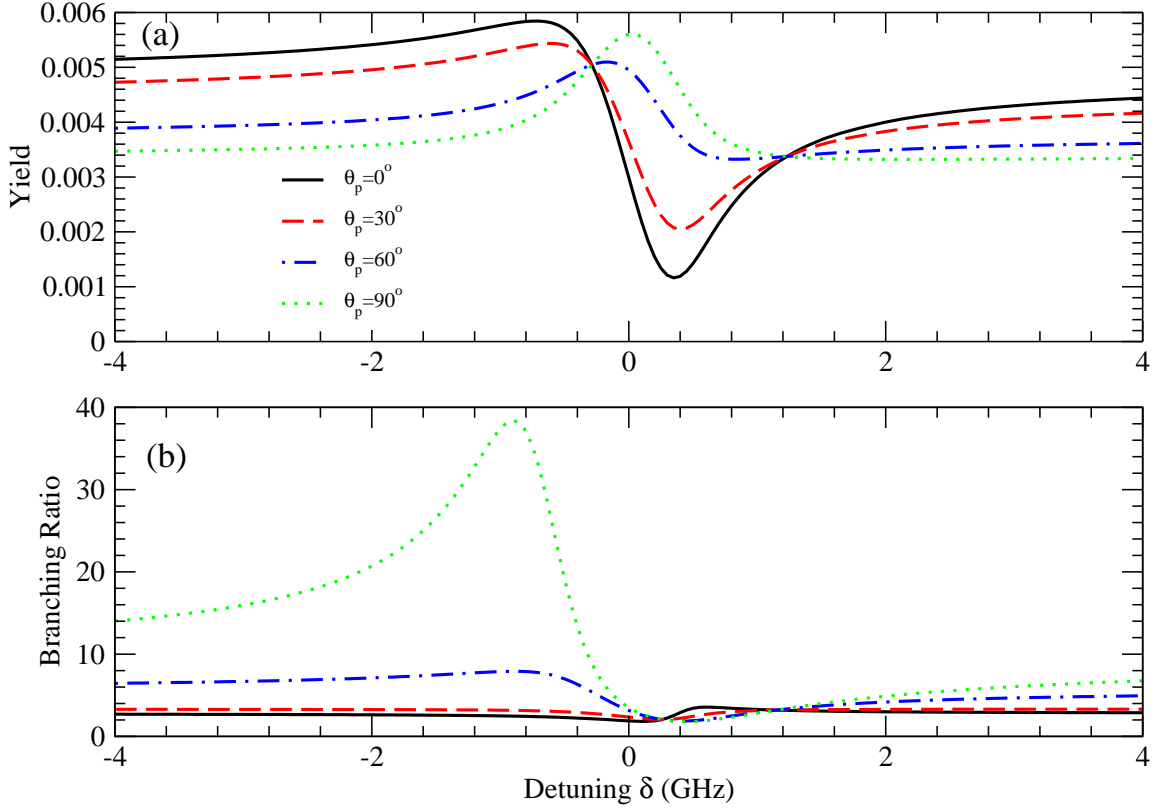}
\caption{Color online. 
(a) Total ionization yield and (b) the branching ratio between the partial 
ionization yields into each $\epsilon s$ and  $\epsilon d$ continuum
for the K $4p_{3/2}-6p_{3/2}$ system as a function of two-photon detuning 
$\delta$.  Pulse durations and peak laser intensities are chosen to be
$\tau_{p} = 1$ ns and $I_{p} = 1 $MW/cm$^{2}$, and $\tau_{d} = 10$ ns 
and $I_{d} = 100$ MW/cm$^{2}$, for the probe and dressing lasers, respectively.
The polarization angle takes the values 
$\theta_p=0^{\circ}, 30^{\circ}, 60^{\circ}$ and $90^{\circ}$.
}
\label{fig7}
\end{figure}

\begin{figure}
\centering
\includegraphics[width=6in,angle=0]{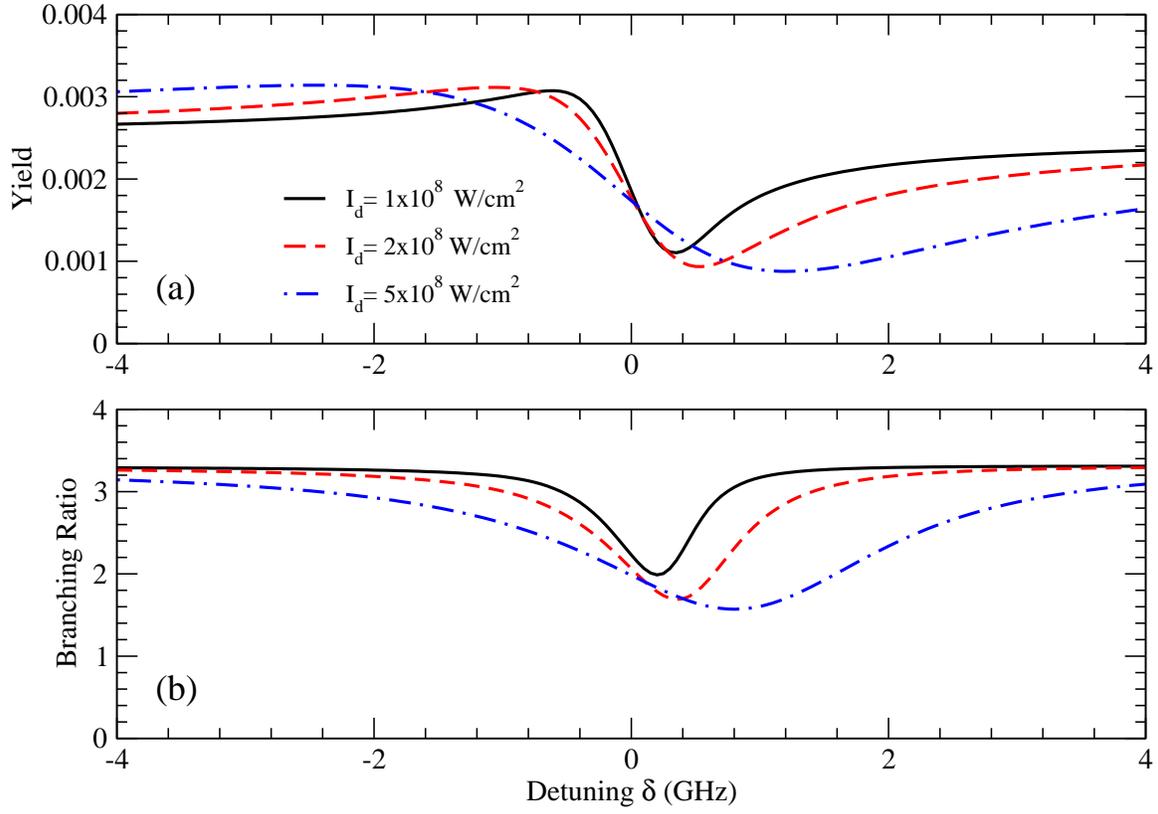}
\caption{Color online. Same as in Fig. \ref{fig4} but for the K $4p_{3/2}-6p_{3/2}$ system.
}
\label{fig8}
\end{figure}

\begin{figure}
\centering
\includegraphics[width=2.0in,angle=0]{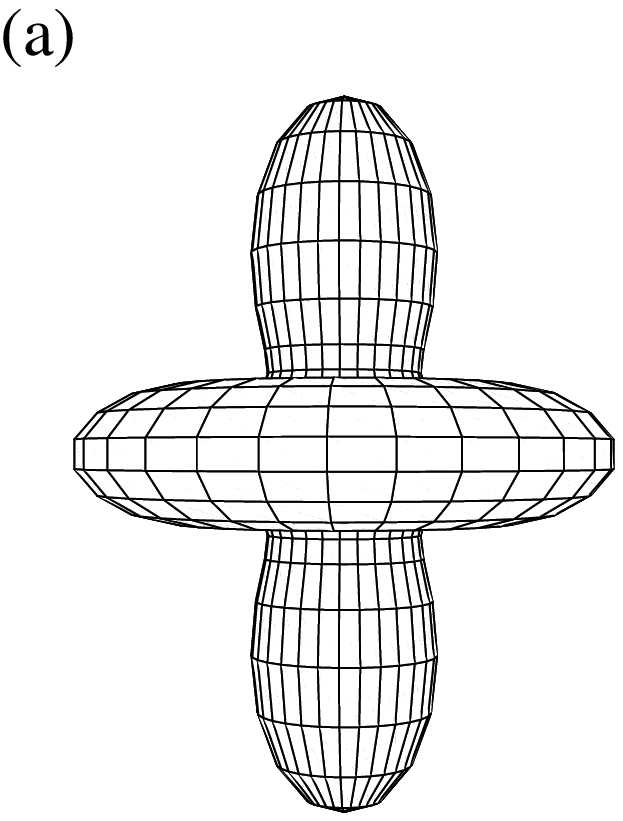}\\
\includegraphics[width=2.0in,angle=0]{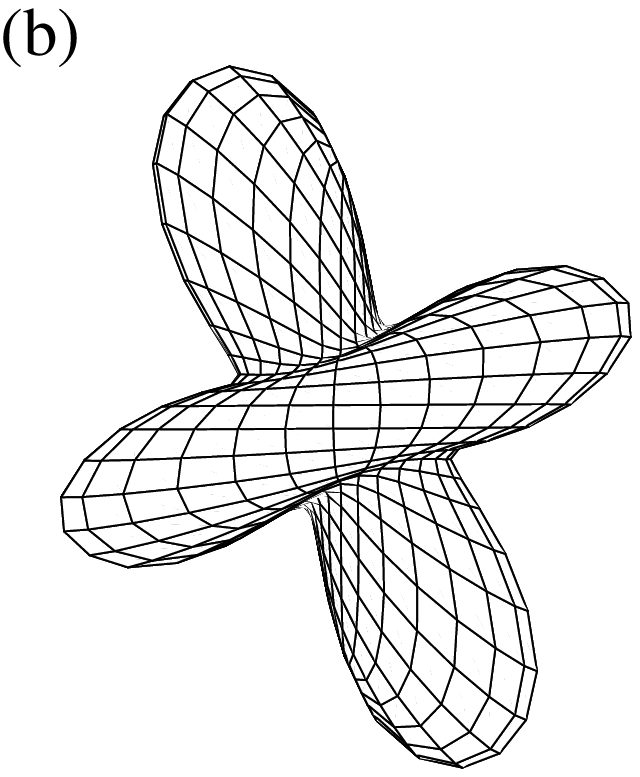}\\
\includegraphics[width=2.0in,angle=0]{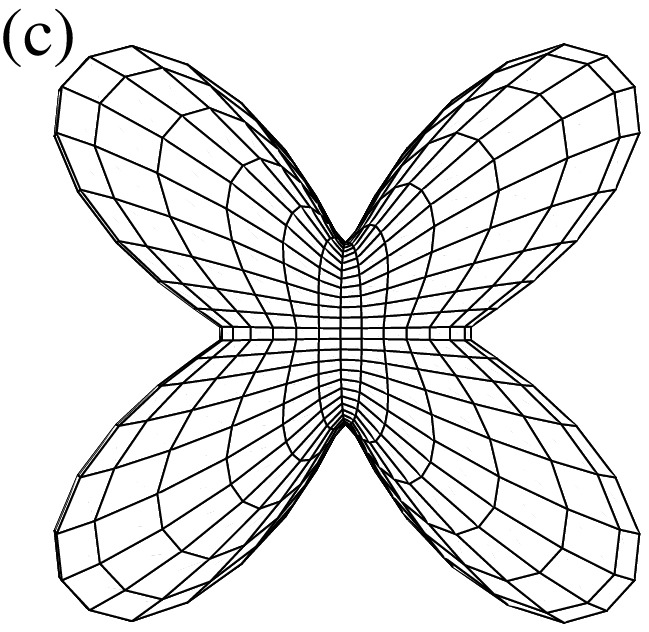}\\
\caption{Same as in Fig. \ref{fig5} but for the K $4p_{3/2}-6p_{3/2}$ system.
 The view point is from the positive $y$-axis.
}
\label{fig9}
\end{figure}

\newpage
\begin{figure}
\centering
\includegraphics[width=2.in,angle=0]{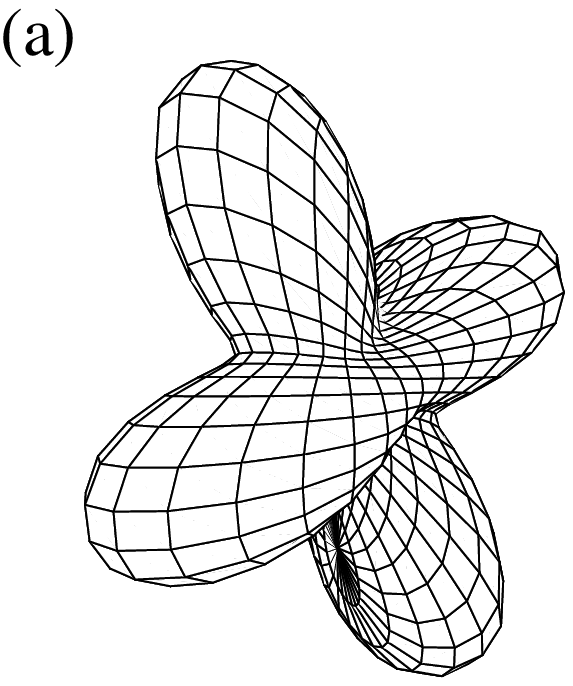}\\
\includegraphics[width=2.in,angle=0]{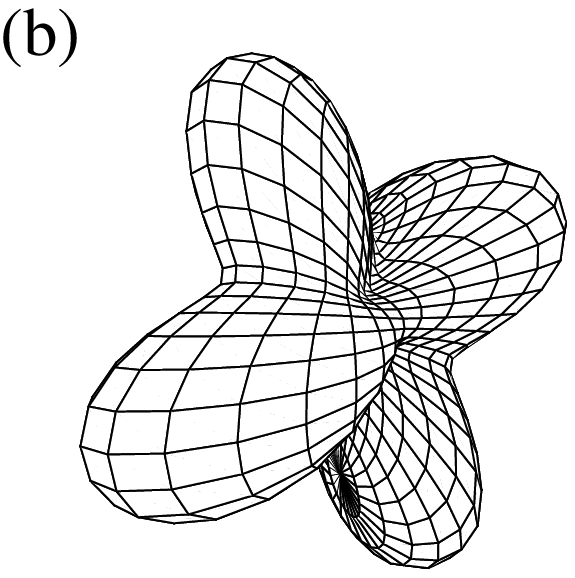}\\
\includegraphics[width=2.in,angle=0]{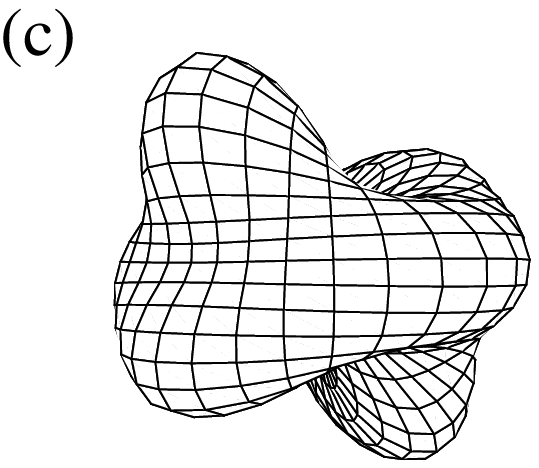}\\
\caption{Three-dimensional photoelectron angular distribution for the 
K $4p_{3/2}-6p_{3/2}$ system at three different two-photon detunings 
$\delta= -4$, $-0.89$, and $0.42$ GHz.  
Pulse durations and peak intensities are $\tau_{p}=1$ ns and 
$I_{p}= 1$ MW/cm$^{2}$ for the probe laser, and $\tau_{d}=10$ ns 
and $I_{p}= 100$ MW/cm$^{2}$ for the dressing laser.
The polarization angle is $\theta_p=60^{\circ}$.}
\label{fig10}
\end{figure}

\begin{figure}
\centering
\includegraphics[width=6in,angle=0]{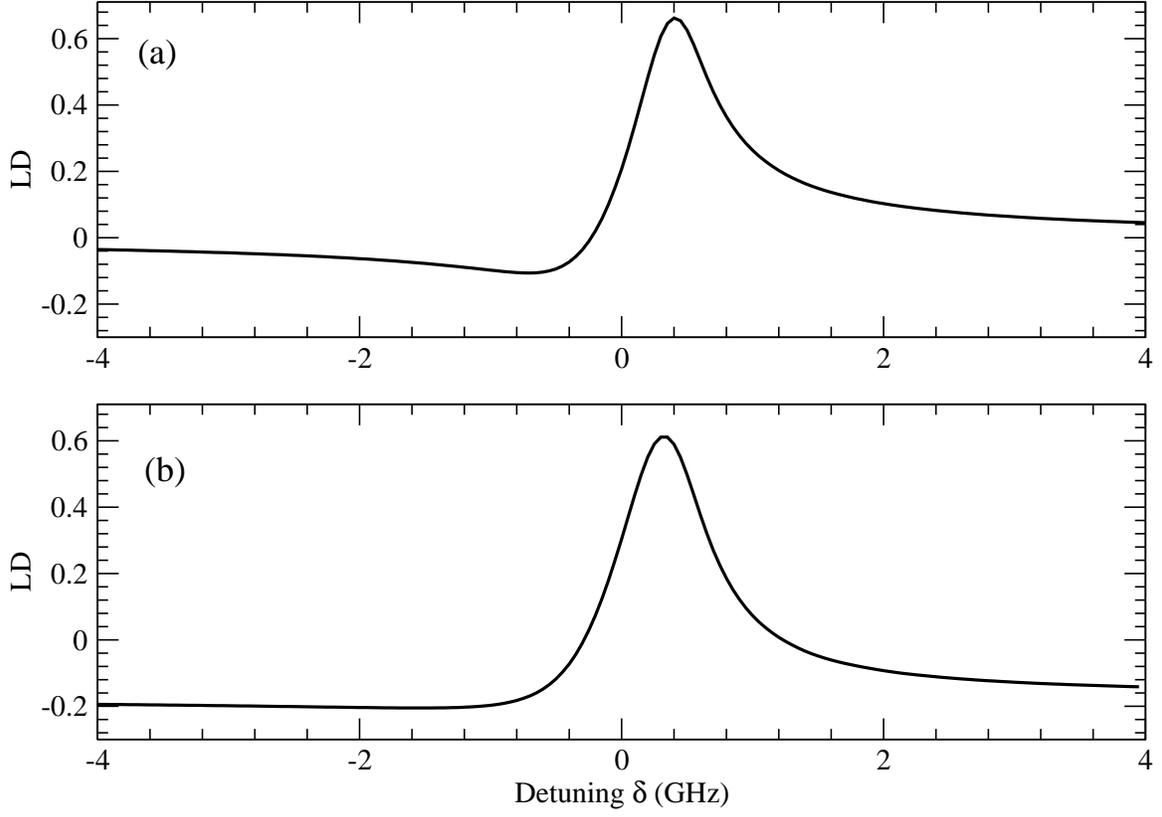}
\caption{Linear dichroism, $LD$, for the (a) K $4p_{1/2}-6p_{1/2}$ system 
and (b) K $4p_{3/2}-6p_{3/2}$ system as a function of two-photon 
detuning $\delta$.  
Pulse durations and peak intensities are $\tau_{p}=1$ ns and 
$I_{p}= 1$ MW/cm$^{2}$ for the probe laser, and $\tau_{d}=10$ ns 
and $I_{p}= 100$ MW/cm$^{2}$ for the dressing laser.
}
\label{fig11}
\end{figure}

\end{document}